\documentclass[a4paper,fleqn,usenatbib]{mnras}

\usepackage{newtxtext,newtxmath}

\usepackage[T1]{fontenc}
\usepackage{ae,aecompl}

\usepackage{graphicx}	
\usepackage{amsmath}	
\usepackage{amssymb}	
\usepackage{float}
\usepackage{multirow}

\usepackage[usenames, dvipsnames]{color}

\def\code#1{\texttt{#1}}

\title[AIMS - Asteroseismic Inference on a Massive Scale]{AIMS - A new tool for stellar parameter determinations using asteroseismic constraints}
\author[B. M. Rendle et al.]
{Ben M. Rendle$^{1,2}$\thanks{E-mail: bmr135@bham.ac.uk},
Ga\"el Buldgen$^{1,2}$,
Andrea Miglio$^{1,2}$,
Daniel Reese$^{3}$,
\newauthor Arlette Noels$^{4}$, Guy R. Davies$^{1,2}$, Tiago L. Campante$^{5,6}$,
William J. Chaplin$^{1,2}$,
\newauthor
Mikkel N. Lund$^{2,1}$, James S. Kuszlewicz$^{7,2}$,
Laura J. A. Scott$^{8}$,
Richard Scuflaire$^{4}$,
\newauthor
Warrick H. Ball$^{1,2}$,
Jiri Smetana$^{9,1}$,
Benard Nsamba$^{5,6}$
\\ \\
$^{1}$School of Physics and Astronomy, University of Birmingham,
Edgbaston, Birmingham, B15 2TT, UK\\
$^{2}$Stellar Astrophysics Centre (SAC), Department of Physics and Astronomy,  Aarhus University, Ny Munkegade  120, DK-8000
Aarhus C,\\ Denmark\\
$^{3}$LESIA,  Observatoire  de  Paris,  PSL  Research  University,  CNRS,  Sorbonne  Universit\'{e}s,  UPMC  Univ.   Paris  06,  Univ.   Paris  Diderot, \\Sorbonne Paris Cit\'{e}, 92195 Meudon, France\\
$^{4}$STAR Institute, University of Li\`{e}ge, Belgium\\
$^{5}$Departamento de F\'{i}sica e Astronomia, Faculdade de Ci\^{e}ncias da Universidade do Porto, Rua do Campo Alegre, s/n, PT4169-007 Porto, \\Portugal\\
$^{6}$ Instituto de Astrof\'{\i}sica e Ci\^{e}ncias do Espa\c{c}o, Universidade do Porto,  Rua das Estrelas, PT4150-762 Porto, Portugal\\
$^{7}$Max-Planck-Institut f\"{u}r Sonnensystemforschung, Justus-von-Liebig-Weg 3, 37077 G\"{o}ttingen\\
$^{8}$Keele Astrophysics Group, School of Chemical and Physical Sciences, Faculty of Natural Sciences, Keele University, ST5 5BG, UK\\
$^{9}$Imperial College London, Department of Physics, South Kensington Campus, London, SW7 2AZ, UK}

\date{Accepted XXX. Received YYY; in original form ZZZ}

\pubyear{2019}

\begin{document}
\label{firstpage}
\pagerange{\pageref{firstpage}--\pageref{lastpage}}
\maketitle

\begin{abstract}

A key aspect in the determination of stellar properties is the comparison of observational constraints with predictions from stellar models. Asteroseismic Inference on a Massive Scale (AIMS) is an open source code that uses Bayesian statistics and a Markov Chain Monte Carlo approach to find a representative set of models that reproduce a given set of classical and asteroseismic constraints. These models are obtained by interpolation on a pre-calculated grid, thereby increasing computational efficiency. We test the accuracy of the different operational modes within AIMS for grids of stellar models computed with the Li\`{e}ge stellar evolution code (main sequence and red giants) and compare the results to those from another asteroseismic analysis pipeline, PARAM. Moreover, using artificial inputs generated from models within the grid (assuming the models to be correct), we focus on the impact on the precision of the code when considering different combinations of observational constraints (individual mode frequencies, period spacings, parallaxes, photospheric constraints,...). Our tests show the absolute limitations of precision on parameter inferences using synthetic data with AIMS, and the consistency of the code with expected parameter uncertainty distributions. Interpolation testing highlights the significance of the underlying physics to the analysis performance of AIMS and provides caution as to the upper limits in parameter step size. All tests demonstrate the flexibility and capability of AIMS as an analysis tool and its potential to perform accurate ensemble analysis with current and future asteroseismic data yields.

\end{abstract}




\section{Introduction} \label{Intro}

At present, asteroseismic supporting space missions in operation and ground-based networks (TESS, \citealt{2015JATIS...1a4003R}; SONG, \citealt{2014RMxAC..45...83A,2017ApJ...836..142G}) or retired missions (CoRoT, \citealt{2006ESASP1306...33B,2016cole.bookD...3C}; \textit{Kepler}, \citealt{2010Sci...327..977B}; K2, \citealt{2014PASP..126..398H}) have generated high quality data for large ensembles of stars. Further missions are also in preparation (PLATO, \citealt{2014ExA....38..249R}). In order to model these stars, we need pipelines that can efficiently compare observations and models. They must be stable, robust and fast to deal with the current volume of data and the subsequent increases expected in the future.

Often, inferred pipeline properties rely on simple scaling relations of the large frequency separation ($\Delta\nu$) and the frequency of maximum oscillation power ($\nu_{\rm{max}}$) \citep{1995A&A...293...87K}. Though quick and simple to use, more robust estimations can be made when, for example, the average large frequency separations from model predicted radial mode frequencies and the use of gravity mode period spacings are considered in the parameter determinations (see \citealt{2017MNRAS.467.1433R,2017ApJS..233...23S}). Though an improvement, these still do not exploit all of the information the individual modes contain, e.g., presence of acoustic glitches \citep{1988IAUS..123..151V,1990LNP...367..283G,2010A&A...520L...6M,2016A&A...591A..99P,2017ApJ...837...47V} and long term internal structure changes from curvature of the large frequency separation \citep{2017A&ARv..25....1H,2012A&A...537A..30M}.

Multiple asteroseismic modelling techniques have been developed with the objective to fully exploit seismic information (\citealt{2004ApJ...600..419G}, \citealt{2005A&A...441..615M}, \citealt{2008MmSAI..79..660B}, \citealt{2009ApJ...699..373M}, \citealt{2012ApJ...749..109G}, PARAM \citep{2006A&A...458..609D,2014MNRAS.445.2758R,2017MNRAS.467.1433R}; see the KAGES \citep{2015MNRAS.452.2127S,2016MNRAS.456.2183D} and LEGACY \citep{2017ApJ...835..172L,2017ApJ...835..173S} projects for further pipelines).

The use of individual mode frequencies as constraints to the analysis, increases significantly both the precision and accuracy of the inferred masses, radii and age for both main sequence (e.g. \citealt{2014A&A...569A..21L} for a recent review, \citealt{2016A&A...592A..14R} for tests using artificial data,  or results based on \textit{Kepler}'s best data sets by \citealt{2017ApJ...835..173S}) and red giant stars \citep{Huber331,2014A&A...562A.109L,2016A&A...591A..99P,2018MNRAS.475..981L}. Improving the use of seismic information will lead to more precise global stellar properties and allow for testing aspects of the micro- and macro-physics which are currently poorly constrained.

We present here the stellar modelling pipeline, AIMS (Asteroseismic Inference on a Massive Scale, \citealt{2016ascl.soft11014R,2018ASSP...49..149L}). AIMS is a pipeline designed to process the measured individual acoustic oscillation frequencies of stars coupled with classical, spectroscopic or interferometric constraints to provide a powerful diagnostic tool for the determination of stellar properties. Much like \cite{2008MmSAI..79..660B}, \cite{2012ApJ...749..109G}, and BASTA \citep{2015MNRAS.452.2127S}, AIMS uses a Bayesian approach. \cite{2008MmSAI..79..660B} implements an on-the fly model calculation with an MCMC algorithm to produce a representative sample of model parameters. This leads to a higher accuracy but at a significant computational cost, whereas the remaining codes use pre-computed grids (faster calculation time). \cite{2012ApJ...749..109G} and BASTA then evaluate probability distribution functions by scanning the grid. Like \cite{2008MmSAI..79..660B}, AIMS also uses an MCMC algorithm, but what is unique is that it is combined with model interpolation. This provides a compromise between accuracy and efficiency.

This paper details the capabilities and potential of AIMS and its applicability within the scientific community. The paper is set out as follows: Section \ref{AIMS} describes the functionality of the code and section \ref{Grid} describes the input grids containing the models used in the analysis. Sections \ref{Interp_test} and \ref{Obs} discuss the results of the various interpolation tests on the grids and the performance of the program in analysing artificial and real data. Finally, a comparison of the performance of AIMS using different combinations of asteroseismic and classical constraints is given in section \ref{obs_cons}. The results of these tests are discussed with a summary of the work in section \ref{Discussion}.

\section{AIMS} \label{AIMS}

AIMS uses Bayesian statistics and a Markov-Chain-Monte-Carlo (MCMC) algorithm (\code{emcee}, \citealt{foreman-mackey_emcee:_2013}) to select models representative of the input data by interpolating in a pre-defined grid. The combination of these techniques allows for an efficient, comprehensive search of the parameter space defined by the grid parameters. User-defined priors and the likelihood function resulting from the input
constraints shape the exploration of the parameter space. AIMS initialises the grid search in the region of a set of models with the highest posterior probability. This increases the efficiency of the parameter space exploration, which in turn helps the MCMC algorithm converge faster.

The program itself has three modes of functionality: binary grid generation; interpolation testing; and stellar parameter characterisation. The performance and capabilities of interpolation mechanism and stellar parameter determination are tested here. Information on the other functions can be found in the supporting documentation\footnote{AIMS Overview:\\
\url{http://bison.ph.bham.ac.uk/spaceinn/aims/version1.3/}}.

To determine stellar parameters in a Bayesian manner, an affine invariant ensemble Markov chain Monte Carlo (MCMC) sampler \citep{2010CAMCS...5...65G} is implemented 
via the \code{Python} package \code{emcee} developed by \cite{foreman-mackey_emcee:_2013}. For a given data file, the user can employ so-called walkers that are initiated in a tightball configuration (optional), uniformly distributing the walkers within a sphere centred on an initial estimation of the most probable grid model. If tightball is not selected, the walkers are initiated through the sampling of model parameter priors. The step number for the walkers can be user defined. Parallel tempering is available with the option to define the number of temperatures, and the MCMC chains can be thinned.

To determine the properties of targets falling between grid points defined by the evolutionary tracks, AIMS uses a two step interpolation procedure of the model parameters:
\begin{enumerate}
\item Linear interpolation in the chosen evolutionary parameter along a track. 
\item Interpolation between tracks.
\end{enumerate}

This method allows for greater control over the evolutionary parameter (prevention of exceeding the boundaries of evolutionary tracks) and attempts to achieve greater accuracy as consecutive models on an evolutionary sequence are not expected to change significantly.
AIMS includes an accuracy test of the interpolation procedure and an additional program is joined to AIMS to visualise these results as a function of the global grid parameters.

The linear interpolation along a track can be modified 
to use various evolutionary parameters. However, only a parameter varying monotonically as a star evolves should be used to prevent any spurious results or unexpected errors within the interpolation. Examples of such variables include the Helium core mass in red giant branch (RGB) stars or the central hydrogen content for main sequence stars (MS).\\

\section{The Grid - CL\'{E}S with LOSC} \label{Grid}

The analysis performed by AIMS is based upon the exploration of a predefined grid of models. In this work, the grid is parametrised by mass (0.75-2.25 M$_{\odot}$,  in 0.02 M$_{\odot}$ increments), initial metallicity ($Z_{\rm{init}}$) and initial hydrogen content ($X_{\rm{init}}$). The range of $X_{\rm{init}}$ and $Z_{\rm{init}}$ values ([Fe/H] values also included for completeness) used can be found in Table \ref{tab:grid_params}. Fig. \ref{fig:CL\'{E}S_HRD} is a Hertzsprung-Russell diagram (HRD) showing the evolutionary tracks calculated for this grid for a given chemical composition. A gap between the MS and sub-giant branch can be observed due to the selection criteria used to split the nominal grid into specific MS and red giant sub grids, which is described in detail later.

\begin{table}
	\centering
	\caption{The values of $X_{\rm{init}}$, $Z_{\rm{init}}$ and [Fe/H] attributed to the CL\'{E}S grid of models.}
	\label{tab:grid_params}
	\begin{tabular}{cccc} 
		\hline
		$X_{\rm{init}}$ & $Z_{\rm{init}}$ & [Fe/H]\\
		\hline
		0.691 & 0.0300 & 0.25\\
		0.716 & 0.0175 & 0.00\\
		0.731 & 0.0100 & -0.25\\
		0.740 & 0.0057 & -0.50\\
		0.745 & 0.0032 & -0.75\\
		\hline
	\end{tabular}
\end{table}

The grid contains the evolutionary tracks of theoretical stellar models and their frequencies. Here, we considered $\sim 38000$ models, but larger grids of up to $\sim 1.5$ million models have been used in the past. The models were computed using the CLES (Code Li\'{e}geois d'\'{E}volution Stellaire, \citealt{scuflaire_CLES_2008}) stellar evolution code and the frequencies were generated using the LOSC (Li\`{e}ge Oscillation Code, \citealt{scuflaire_liege_2008}) pulsation code. We use the \cite{1993oee..conf...15G} abundances, nuclear reaction rates of \cite{2011RvMP...83..195A}, opacities of \cite{1996ApJ...464..943I} and the FreeEOS equation of state \citep{2012ascl.soft11002I}. The mixing-length parameter was kept to a solar calibrated value of $1.67$ and a convective overshoot of $0.05$ times the local pressure scale height was used, assuming instantaneous chemical mixing and the radiative temperature gradient in the overshooting region. Microscopic diffusion was not included in the grid. The border of the convective zones was calculated following the guidelines of \cite{gabriel_proper_2014} to avoid spurious solutions for the evolution of convective cores.

\begin{figure}
	\includegraphics[width=\columnwidth]{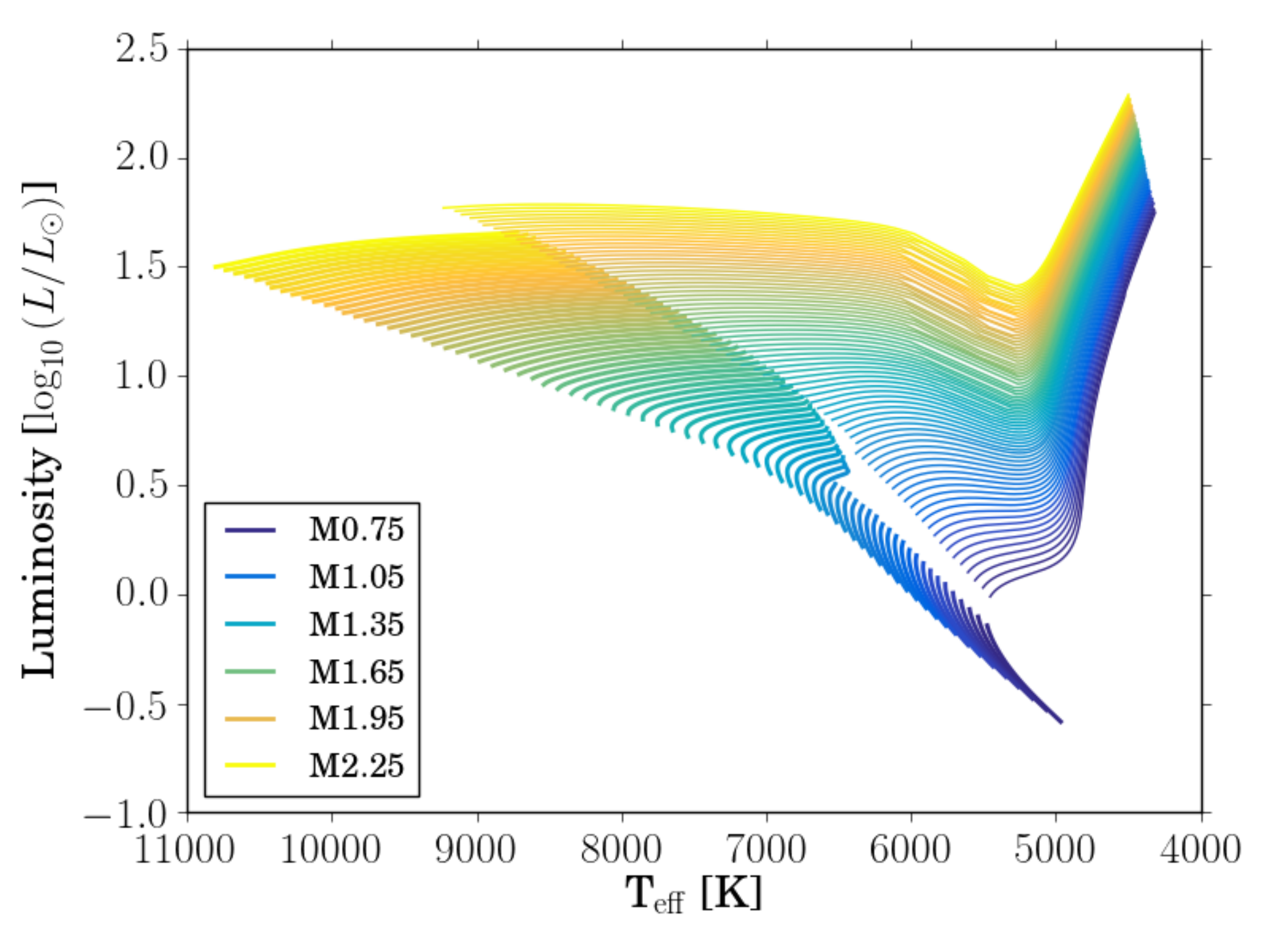}
    \caption{Hertzsprung Russell Diagram displaying the evolutionary tracks found within the CL\'{E}S grid ($X_{\rm{init}}$ = 0.731, $Z_{\rm{init}}$ = 0.0100). The gap between the end of the MS and beginning of the sub-giant branch is due to the Helium core mass fraction selection criterion for the MS and RGB grids.}
    \label{fig:CL\'{E}S_HRD}
\end{figure}

In this work, we used two sub-grids: one for MS and another for RGB stars. We based our criteria on the changes in chemical composition (variations of central hydrogen for the MS, helium core mass for the RGB), effective temperature and $\nu_{\rm{max}}$ values. While AIMS is very versatile in the grids it can use, it should be noted that the tracks must contain a sufficient number of models to ensure an accurate interpolation. On the MS, we included modes with angular degree ($\ell$) values of 0, 1 and 2 whereas the RGB grid only used radial modes ($\ell=0$). This difference stems from intrinsic limitations of AIMS in processing non-radial modes of RGB stars which are highly non-linear. Both grids included radial orders of the frequencies in the range $n = 0-30$. It should be noted that the grids were built to test the functionality of the code that we will describe in sections \ref{Interp_test} and \ref{Obs}.

\section{Interpolation Testing} \label{Interp_test}

The objective of AIMS is to carry out precise asteroseismic analyses. Hence, it is paramount to ensure an accurate interpolation of the determined stellar properties to ensure the reliability of the modelling results. Here, we briefly present the interpolation procedure used in AIMS and the tests that can be made to certify accurate and reliable results.

\subsection{Interpolation Procedure} \label{Interp_Pro}

AIMS uses a two step interpolation process to explore the regions between models, namely:
\begin{enumerate}
\item interpolation between evolutionary tracks
\item interpolation along an evolutionary track
\end{enumerate}

Interpolation between the tracks relies on a multi-dimensional Delaunay tessellation (see \citealt{field_generic_1991} and references therein) of the grid parameters excluding age. The tessellation and subsequent interpolation are carried out by python's \code{scipy.spatial.Delaunay} module which is based on the \code{Qhull}\footnote{\url{http://www.qhull.org/}} package \citep{Barber:1996:QAC:235815.235821}. Using a tessellation approach offers two advantages: the grid does not need to be structured, and fewer tracks (namely $\rm{n}_{\rm{dim}}+1$ as opposed to $2^{\rm{n}_{\rm{dim}}}$, where $\rm{n}_{\rm{dim}} \geq 2$ is the number of dimensions excluding age) are used when interpolating at a given point, accelerating the calculations. During the tessellation, the parameter space is divided into simplices (i.e. triangles in the 2D case, tetrahedra in the 3D case, etc.). For a given point in this space, AIMS searches for the simplex containing it and carries out a linear combination of its vertices (or nodes). The interpolation coefficients correspond to barycentric coordinates provided by \code{scipy.spatial.Delaunay}. These coefficients are simply the ratios between the volumes of the reduced simplices where one of the vertices has been replaced by the point where the interpolation is carried out and the volume of the original simplex. 

\begin{figure}
\includegraphics[width=\columnwidth]{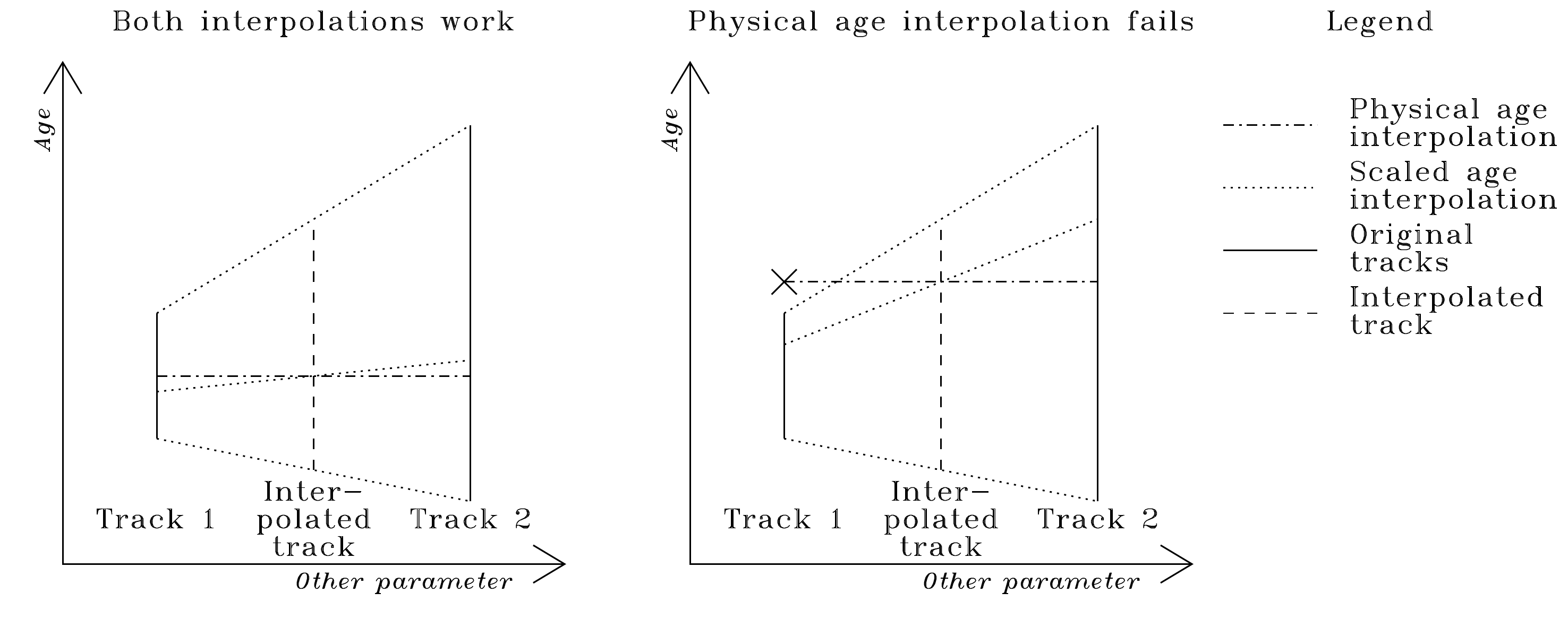}
\caption{Comparison of physical and scaled age interpolation. In the right panel, only scaled age interpolation works. \label{fig:age_interpolation}}
\end{figure}

Interpolation along the tracks consists of a linear interpolation in age between the two closest models.  Points outside the tracks are rejected, i.e. AIMS does not perform extrapolation. AIMS can either interpolate according to the physical age, or according to an age parameter which has been scaled to go from 0 to 1 along the track (e.g. helium core mass in red giants). This latter option is more robust as it is less likely to lead to extrapolation (and hence model rejection) when the two interpolation steps are combined. Indeed, the point where the interpolation is being carried out only needs to be within the age span of the interpolated track rather than having to lie within the age span of all tracks involved in the interpolation, as illustrated in Fig.~\ref{fig:age_interpolation}.

\begin{figure}
\includegraphics[width=\columnwidth]{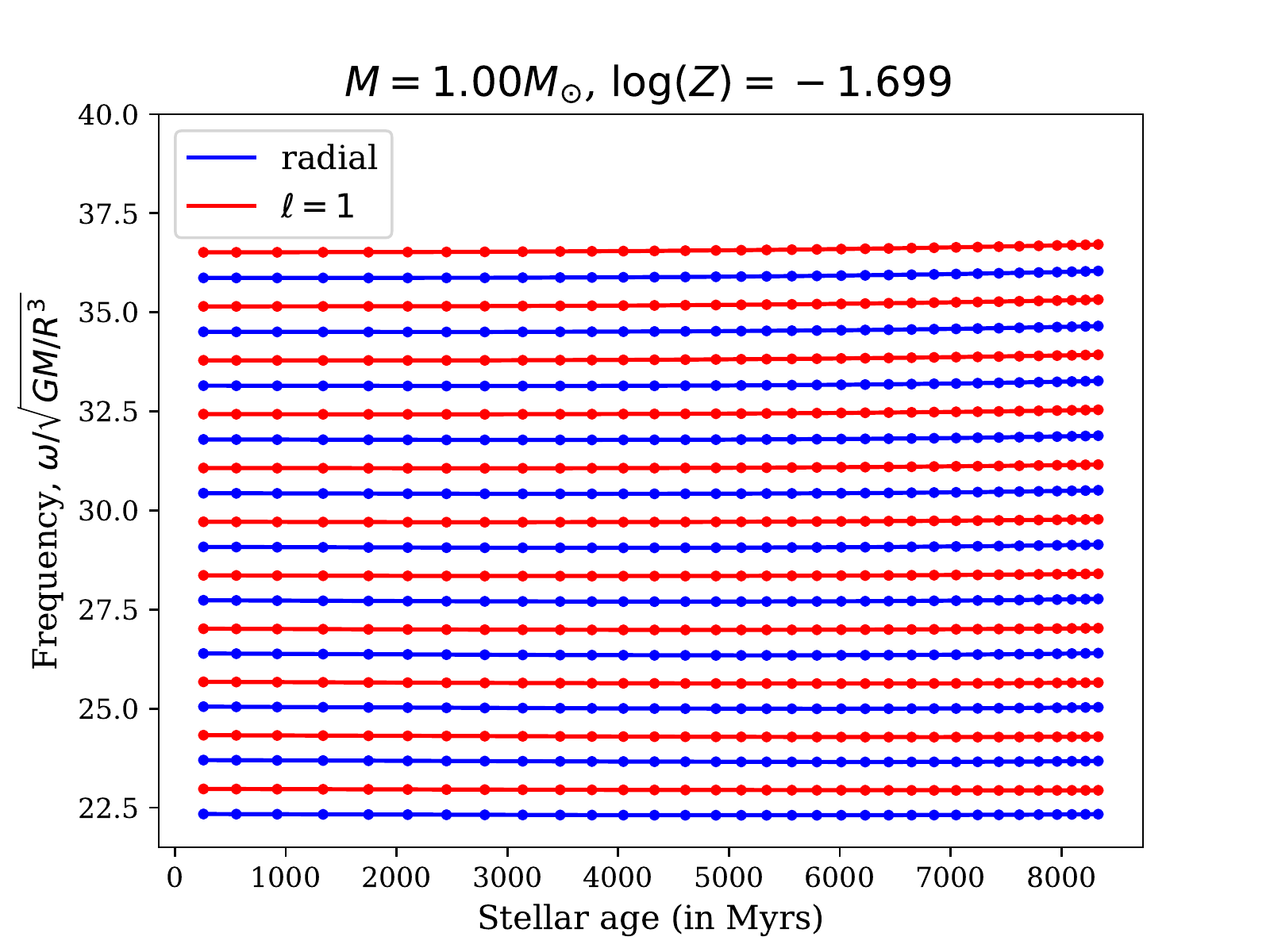}
\includegraphics[width=\columnwidth]{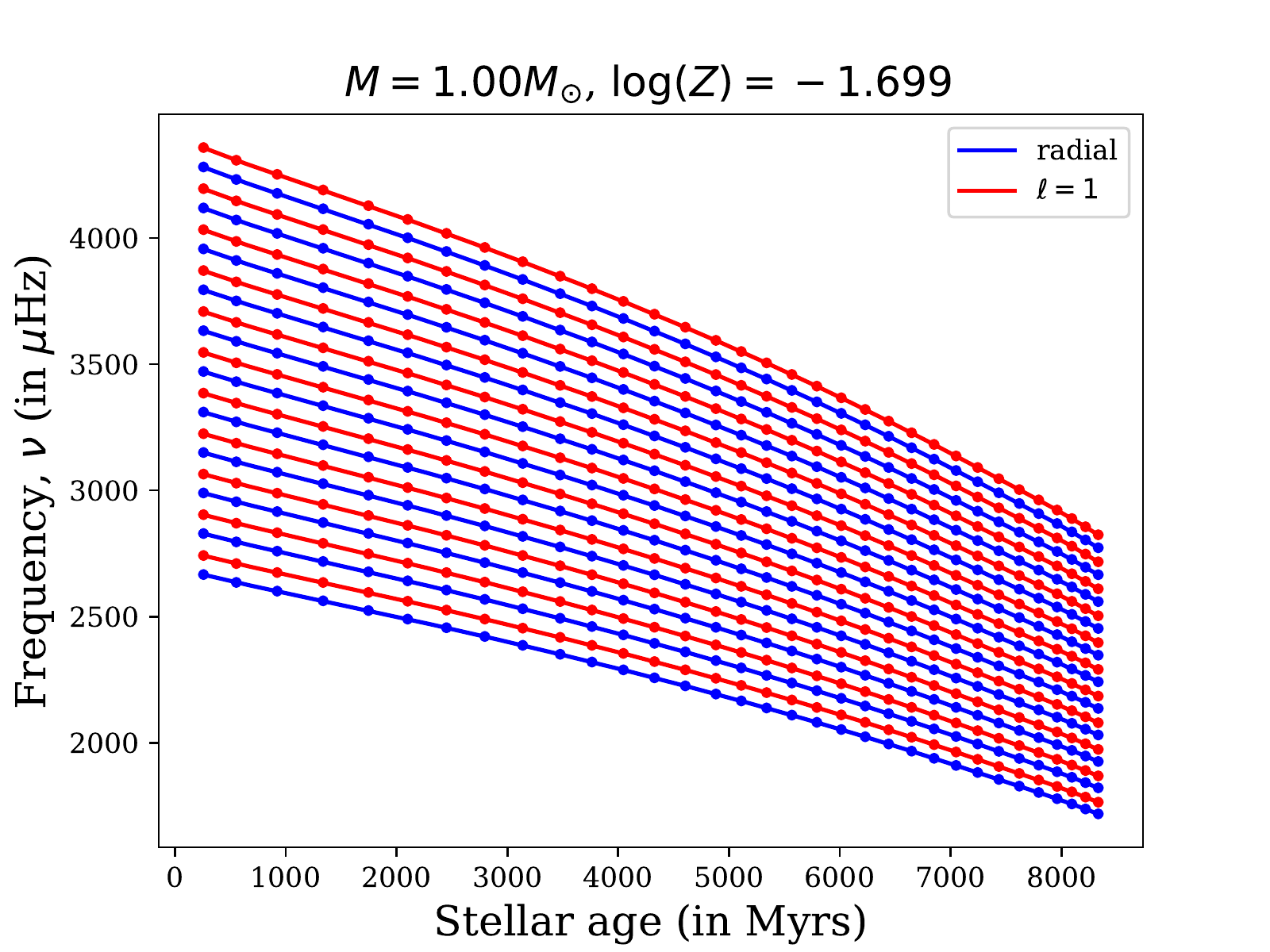}
\caption{Frequencies as a function of stellar age along an evolutionary track.  The upper panel corresponds to non-dimensional frequencies and the lower panel to their dimensional counterparts.  The symbols correspond frequencies from the non-interpolated models whereas the continuous lines represent the interpolated frequencies. \label{fig:frequency_interpolation}}
\end{figure}

The determined coefficients are then used to interpolate the models by linearly combining the global parameters $M$, $X_0$ (the initial hydrogen content), $Z_0$ (the original metallicity), $T_{\mathrm{eff}}$, and $\rho$ (the mean density). The radius and luminosity are then determined self-consistently from these interpolated parameters using the relations:
\begin{equation}
R = \left(\frac{3M}{4\pi \rho}\right)^{1/3}, \qquad
L = 4\pi \sigma R^2 T_{\mathrm{eff}}^4.
\end{equation}
We note that even the constant $\sigma$ is interpolated linearly in case there were any departures from the Stefan-Boltzmann law. Any supplementary user-provided global parameters are interpolated linearly.  The mean density is interpolated linearly rather than the radius in order to be consistent with the results from \code{InterpolateModel}.\footnote{\url{https://bison.ph.bham.ac.uk/spaceinn/interpolatemodel/}, a program which interpolates the acoustic structure of models using outputs from AIMS.} Non-dimensional frequencies, $\omega/\sqrt{GM/R^3}$, with the same $n$ and $\ell$ identification are interpolated linearly rather than their dimensional counterparts, as they vary much more slowly as a function of stellar parameters, as illustrated in Fig.~\ref{fig:frequency_interpolation}. They are subsequently multiplied by $\sqrt{GM/R^3}$, using the interpolated values of $M$ and $R$, in order to remain consistent with the interpolated global parameters. The interested reader is referred to the AIMS documentation for additional information.

\subsection{Interpolation Results} \label{Interp_Res}

In this section, we present the tests included in AIMS to check the suitability of the interpolation procedure to fit observational data. We compare the interpolation errors to the typical uncertainties of observed targets found in the literature. On the MS, we used 16-Cyg A, which yields a median frequency uncertainty on the $l=0$ modes of 0.08 $\mu$Hz (-1.097 in log$_{10}$) and a smallest uncertainty of 0.04 $\mu$Hz (-1.398 in log$_{10}$). On the RGB, we use KIC4448777, which has a median frequency uncertainty on the $l=0$ modes of 0.018 $\mu$Hz (-1.745 in log$_{10}$) and a smallest uncertainty of 0.014 $\mu$Hz (-1.854 in log$_{10}$).

\subsection{Interpolation along evolutionary tracks} \label{TI_age}

\begin{figure*}
\includegraphics[width=\textwidth]{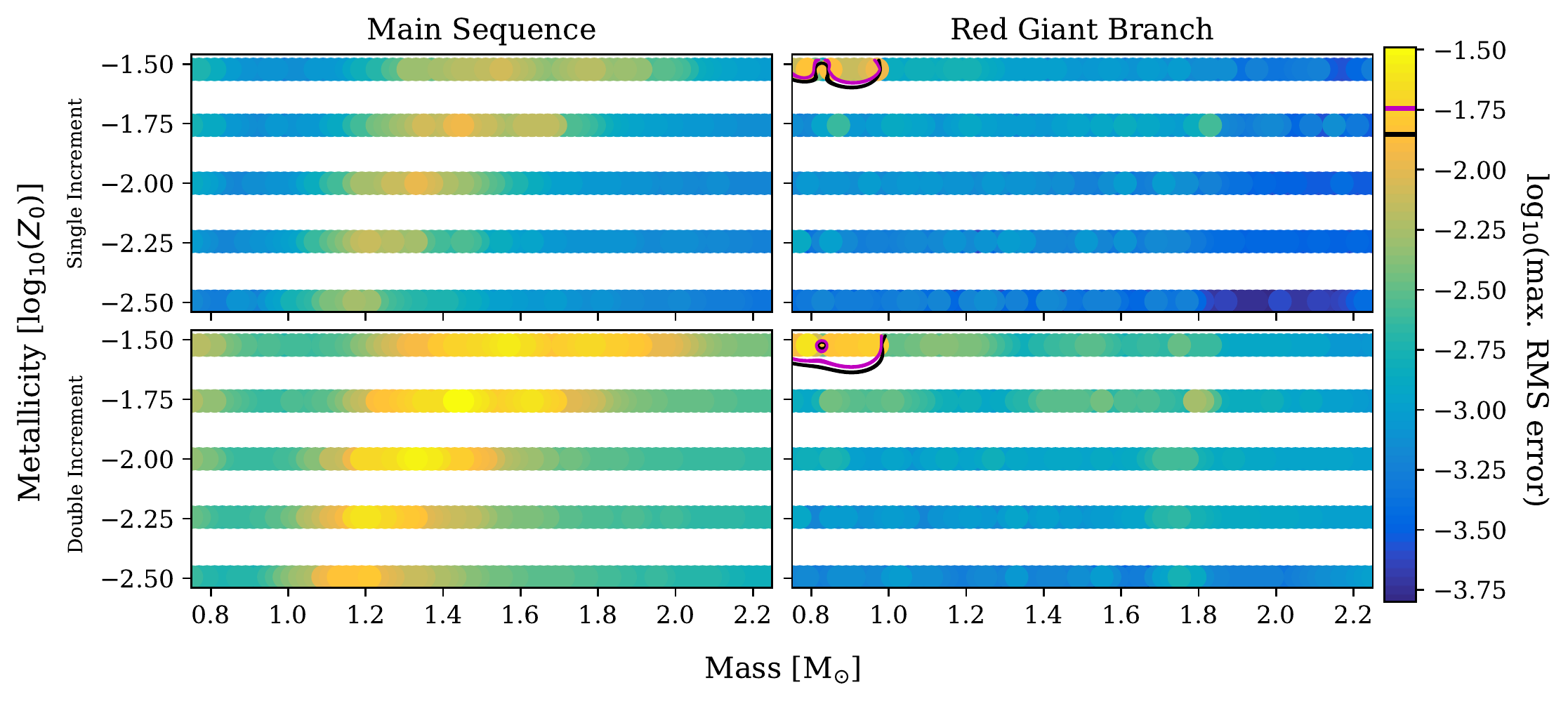}
\caption{Average frequency interpolation errors over the range $\nu_{\mathrm{max}} \pm 0.2 \nu_{\mathrm{max}}$ along evolutionary tracks for radial modes from the MS grid (left panels) and the RGB grid (right panels). The upper panels use increments of $1$ along the tracks whereas the lower panel corresponds to increments of $2$. The magenta and black contours correspond to the average and smallest error bars of KIC4448777. The average and lowest uncertainties for 16-Cyg A are not shown as they are greater than the uncertainty range shown}
\label{fig:interp_TI_age}
\end{figure*}

The evaluation of the interpolation errors along an evolutionary track is made by testing how well both frequencies and global parameters of each model can be recovered from adjacent models at 1 and 2 increments away. Figure~\ref{fig:interp_TI_age} shows the RMS average interpolation errors on the frequencies over the range $\nu_{\mathrm{max}} \pm 0.2 \nu_{\mathrm{max}}$ for the MS and RGB grids detailed in Section \ref{Grid}. Overall the errors are smaller than the smallest frequency uncertainty of 16-Cyg A over the tested frequency range for both single and double increments. The behaviour of the interpolation error is in line with the expectations for a simple linear interpolation, as it increases by a factor of $\sim 4$. Increased errors are seen between 1.2 and 1.8 $M_{\odot}$ and are linked to the onset of a convective core during the evolution. The results are, however, satisfactory as they are well below the observational error bars.

On the RGB, the interpolation errors remain below the smallest and average uncertainties for KIC4448777 apart from a small region at low masses and high metallicities, which represents 2 to $3\,\%$ of the models, as highlighted in the right panels of Fig.~\ref{fig:interp_TI_age} by the black and magenta contours respectively. Again, using double increments in the interpolation leads to an increase in line with numerical expectations. While the RGB results may seem worse than for the MS, one must bear in mind the comparatively smaller error thresholds on the RGB. The RGB interpolation errors remain actually smaller than the MS, as shown in Fig.~\ref{fig:interp_TI_age} and can of course be reduced by refining the grid.

\subsection{Cross Track Interpolation} \label{TI}

As a result of the multi-dimensional character of the parameter space and the use of Delaunay tessellation, the approach used to test cross-track interpolation in AIMS is quite different. The grid is partitioned in two sub grids: one to form the simplices for the interpolation and one containing the tracks to be recovered via interpolation. The partition is made randomly to avoid biasing the test towards one of the directions. This, however, means that the models are not always adjacent to the interpolated ones, reducing the representativity with respect to what is done in practice.

\begin{figure*}
\includegraphics[width=\textwidth]{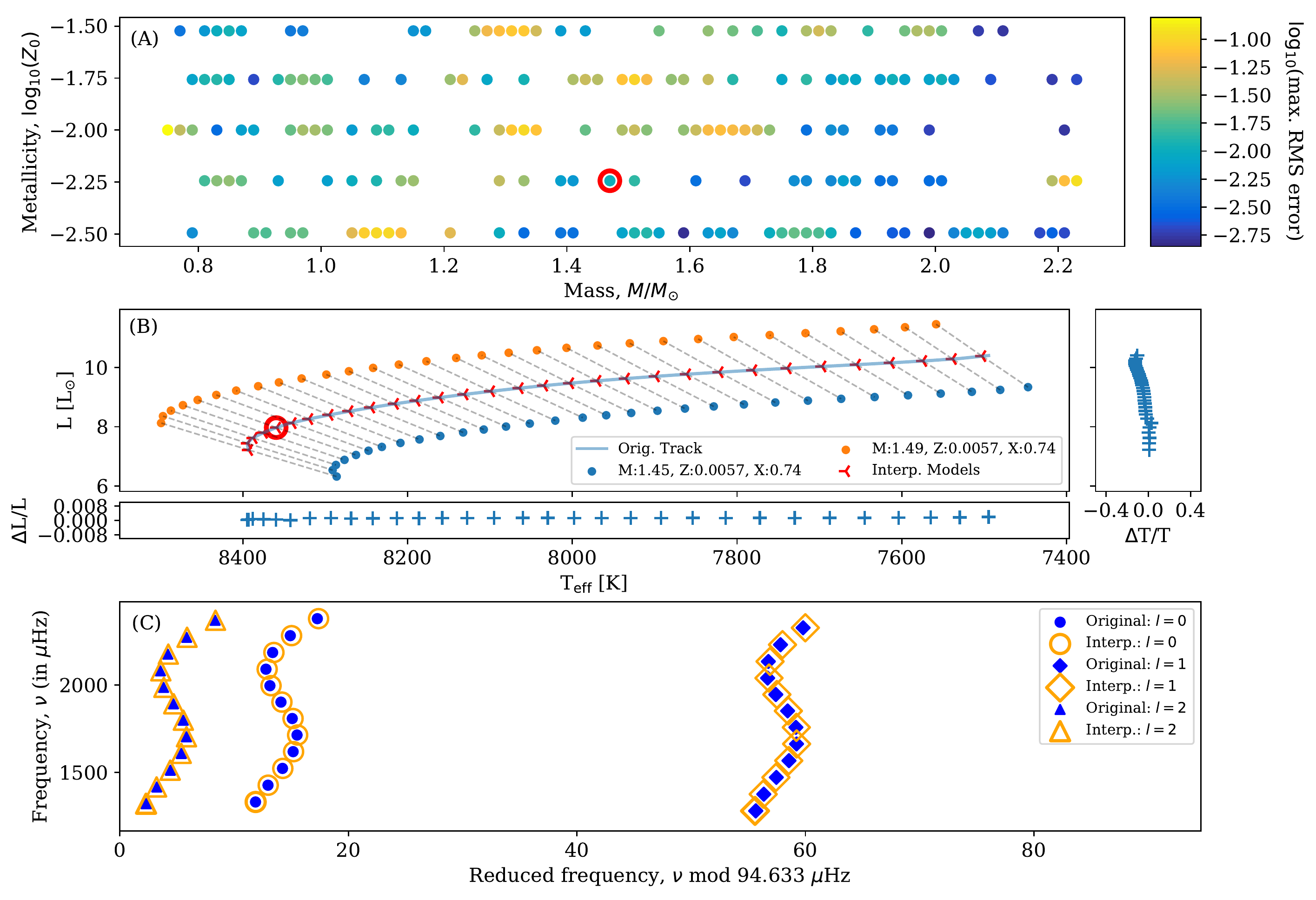}
\caption{An example of results achieved recovering a 1.47 M$_{\odot}$, $X_{\rm{init}}$ = 0.740, $Z_{\rm{init}}$ = 0.0057 track during interpolation testing: 
(A) - Colour map of the maximum interpolated frequency uncertainty along each track. Points represent the positions re-interpolated tracks; the red circle shows the position of the track used in part (B). This track has a maximum log$_{10}$ uncertainty on the interpolated frequencies of -1.990. The red circle highlights the location of the track. Grey points and lines show the triangulation simplices for the interpolation. Red diamonds donte tracks not interpolated due to no triangulation being possible.
(B) - Hertzsprung-Russell Diagram showing the original track (blue line), interpolated track (red markers) and the models the track was interpolated from. Models used for interpolation are connected to the respective interpolated models by grey dashed lines and are shifted by 0.5 L$_{\odot}$ for additional clarity. The fractional difference residuals in luminosity and $T_{\rm{eff}}$ between the original and interpolated models are shown. The $T_{\rm{eff}}$ residuals have been inflated by a factor of 100. The red circle marks the model used in (C).
(C) - An echelle diagram showing the original (blue, closed) and interpolated (orange, open) frequencies for the highlighted model in (B). Full frequency range is shown with diagram modulated by the original model $\Delta\nu$ value. All frequencies have been shifted by 5$\mu$Hz in the x-direction for clarity.}
\label{fig:MS_trackI} \end{figure*}

Panel (A) of Fig. \ref{fig:MS_trackI} displays the recovered sub-grid from the MS interpolation. The RMS average interpolation errors are consistent with the MS and RGB values for along track interpolation, but extend to higher values in some regions. These predominately follow the increased error pattern in Fig. \ref{fig:interp_TI_age}. Higher uncertainties are expected though, as a greater range of parameter space than normally used is interpolated across. The maximum interpolation errors are the order of the average frequency uncertainty of 16-Cyg A. The errors are acceptable as the values are consistent with average observation uncertainties for interpolations over greater ranges than will be executed during real parameter determination.

Selecting a model from the recovered sub-grid, one can see how well the interpolation has reconstructed the original track. Panel (B) of Fig. \ref{fig:MS_trackI} shows the recovered 1.47 M$_{\odot}$, $X_{\rm{init}}$ = 0.740, $Z_{\rm{init}}$ = 0.0057 track and panel (C) an echelle diagram for the original and interpolated frequencies for a single model. The interpolated temperatures, luminosities and frequencies vary fractionally about the original values, illustrating further the accuracy of the interpolation method.

Appendix \ref{Track_Interp} shows an example of the RGB grid. The variation in the residuals is minimal, confirming the proper behaviour of the interpolation. In additional tests, some instances show variations from the expected values along sections of the track. These features are largest when interpolating between grid points separated widely in mass (> 0.05 M$_{\odot}$) or metallicity, consistent with the regions of increased uncertainty in Fig. \ref{fig:MS_trackI} outside of the convective onset region.

\subsection{Other Parameters} \label{Other_Params}

These tests can also be performed for parameters such as mass, radius, luminosity, effective temperature and surface metallicity ratios. The results are not presented here to avoid redundancy but details an be found in Appendix \ref{param_interp}. Again, these tests validated the quality of the grid at both the single and double increment level.

\section{Observational Outputs and Constraints} \label{Obs}

In this section, we present the robustness and accuracy of AIMS in reproducing accurately and precisely stellar parameters. The results presented here illustrate the absolute precision AIMS could achieve for the specific grid used in this study. It should be noted that the performance will depend on the grid and the free parameters included.

\subsection{Artificial Data} \label{Art_Data}

At first, tests were performed using models from the underlying CL\'{E}S grids. An observation file for a single, randomly selected model containing the artificial frequencies, $T_{\rm{eff}}$, $\nu_{\rm{max}}$, luminosity ($L$) and [Fe/H] values for the track was generated. This track was then removed from the grid. The input file was perturbed 100 times to simulate noise in the data signal. This artificial target was fitted using 100 AIMS runs and the average values from these consecutive fits and their uncertainties were used to determine the degree of success of the procedure.

The asteroseismic constraints selected for use in the analysis were the individual mode frequencies. There are multiple options that can be selected for the seismic constraints, with each having a slightly different effect on the output parameters.
Other constraints such as the average $\Delta\nu$ and various frequency separation ratios (r$_{0,1}$, r$_{0,2}$, r$_{1,0}$) could have been used. Using individual mode frequencies gives the smallest uncertainties on the derived parameters, but the final parameter values remain consistent throughout. 

One should be cautious though as individual frequencies are not individually unique constraints and can lead to an underestimation of uncertainties. They are also significantly affected by surface effects (this is true of other parameters, e.g. mass, but the changes are more obvious in such cases), at a level such that the precision of the fit is determined by the uncertainties in the surface correction rather than the frequencies (see \citealt{2018arXiv180808391B} for examples).

\subsubsection{Main Sequence} \label{MS_art}

\begin{figure}
	\includegraphics[width=\columnwidth]{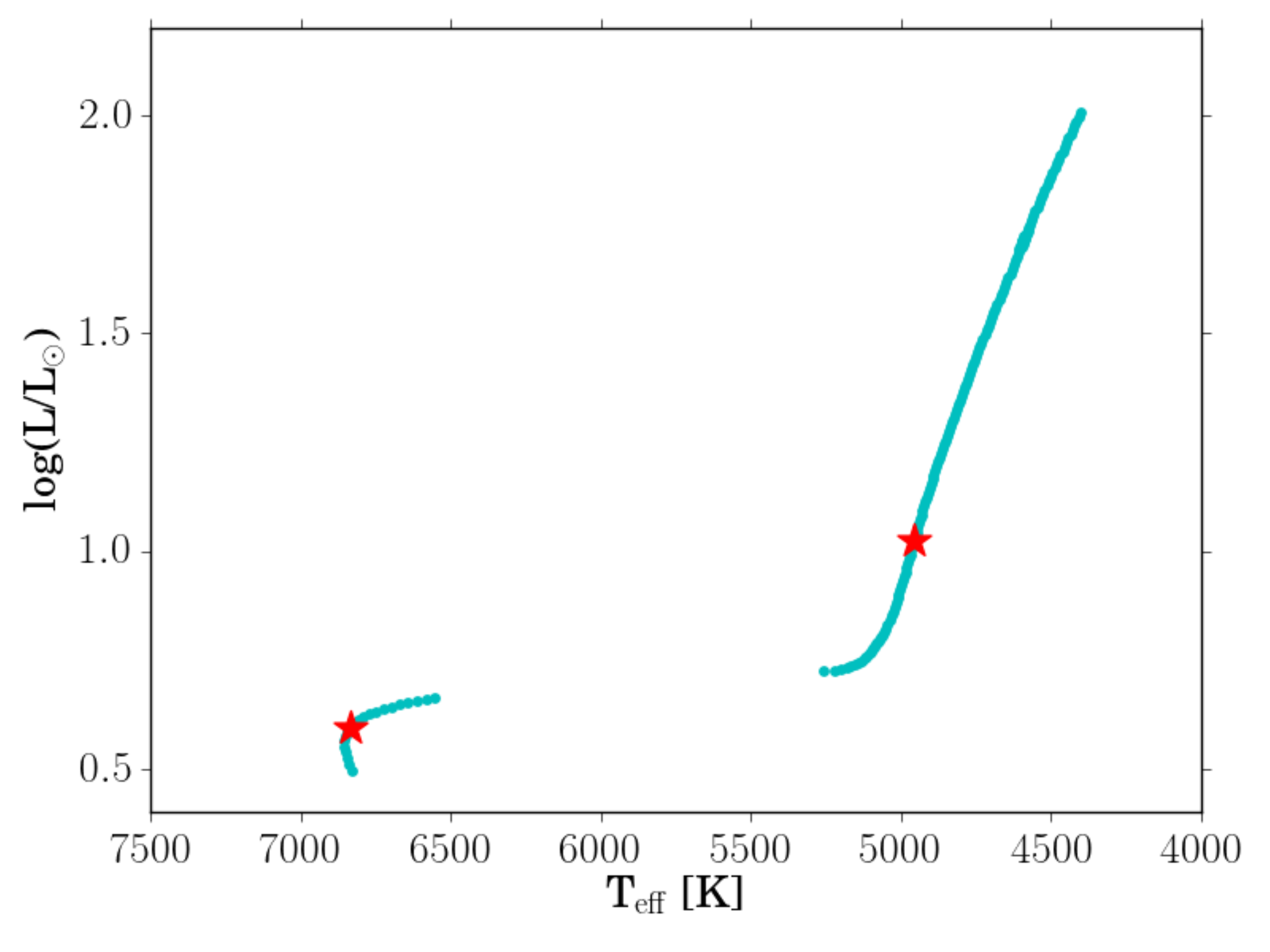}
    \caption{HR diagram showing the evolution of the 1.27 M$_{\odot}$, $Z_{\rm{init}} =$ 0.0100, $X_{\rm{init}} =$ 0.731 track. The red stars indicate the positions of the models selected for the artificial data analysis on the MS and RGB. Models prior to the zero-age-main-sequence (ZAMS) have been removed for clarity and final grid selection criteria have been applied.}
    \label{fig:M1.27_evo}
\end{figure}

A 1.27 M$_{\odot}$, $Z_{\rm{init}}=$ 0.01 and $X_{\rm{init}} =$ 0.731 MS model with 21 mode frequencies (7 of each of $l = 0,1,2$) was selected (see Fig. \ref{fig:M1.27_evo}).
We used the uncertainty distribution of 16-CygA \citep{2015MNRAS.446.2959D} for our artificial target. The magnitude of the uncertainties are of the same order as those used in the ``Sun-as-a-star'' tests in section \ref{Real}. Uncertainties in [Fe/H] and $T_{\rm{eff}}$ were of order 0.1 dex and 80 K respectively. The uncertainty on the luminosity was selected to be of order 3$\%$ based on Gaia \citep{2018A&A...616A...1G} parallaxes, with a large proportion of the uncertainty due to the applied bolometric corrections \citep{torres_use_2010,casagrande_synthetic_2014,casagrande_synthetic_2018}. No surface effects were used for both the artificial target and the seismic modelling.

\begin{figure*}
	\begin{tabular}{cc}
	\includegraphics[width=0.48\textwidth,keepaspectratio]{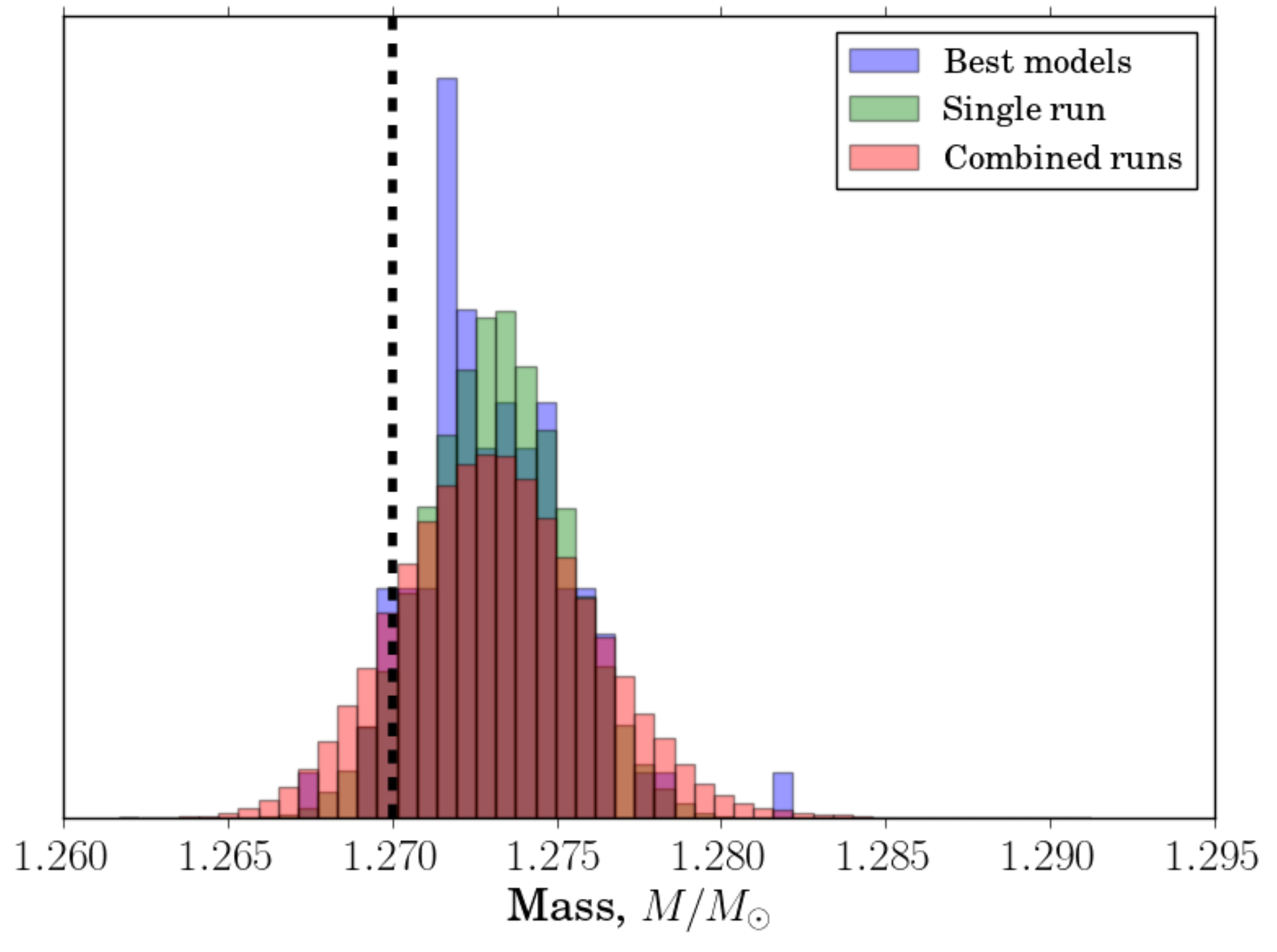} &
    \includegraphics[width=0.48\textwidth,keepaspectratio]{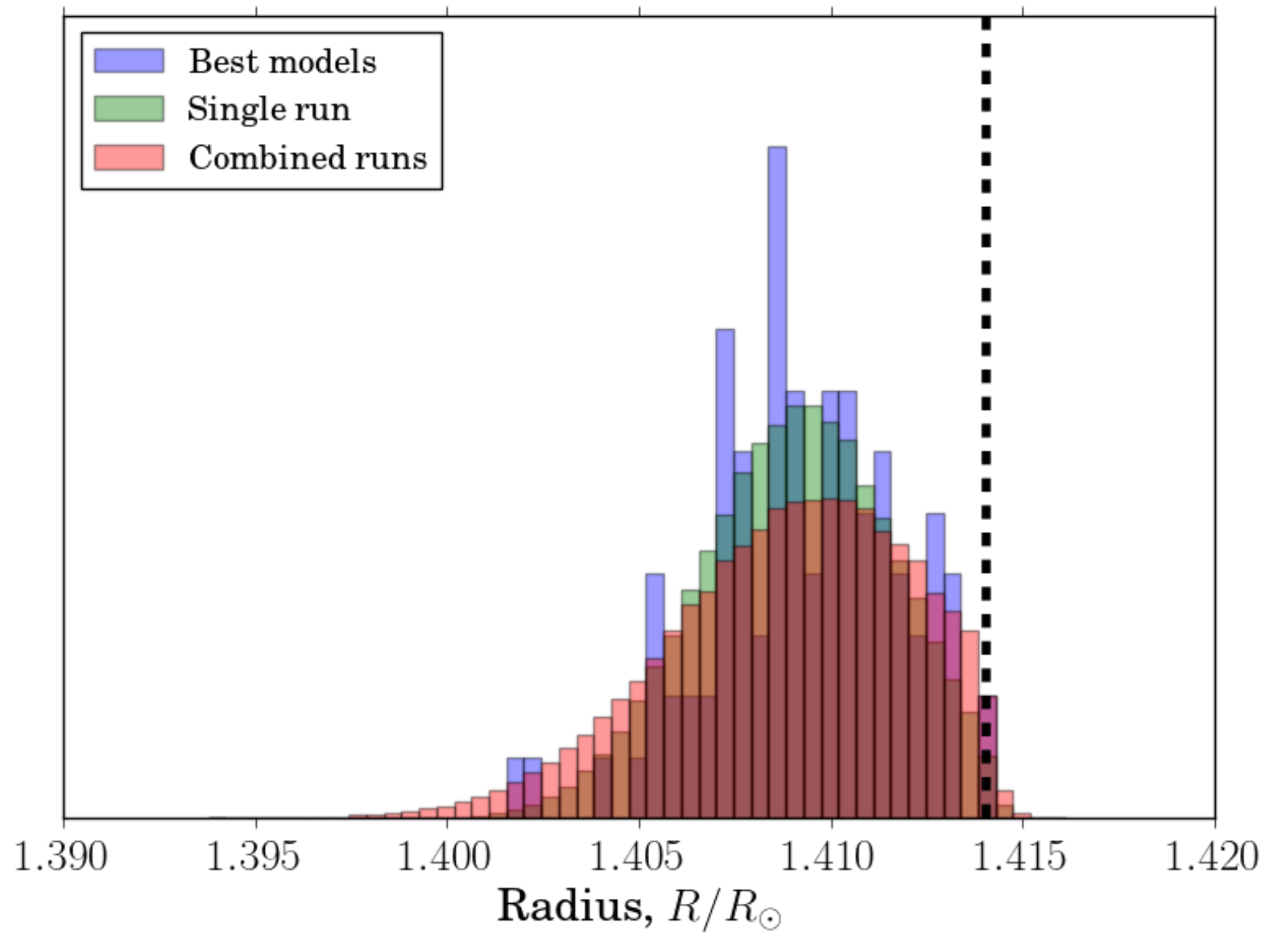}\\
    \end{tabular}
    \caption{The mass (left) and radius (right) PDF distributions for RGB the single model (green), 100 realisations (red) and best models from the MCMC runs (blue). The model mass and radii are 1.27 M$_{\odot}$ and 1.414 R$_{\odot}$, indicated by the vertical black, dotted line.}
    \label{fig:hist_rad}
\end{figure*}

The values and uncertainties of the unperturbed model, the 100 realisations (combined runs) and the best models determined by the MCMC process for each of the perturbed runs were compared to the real values of the model. Examples of the PDF distributions for the mass and radius for each of the 3 trials are shown in Fig. \ref{fig:hist_rad}.

The evidence from Fig. \ref{fig:hist_rad} indicates that the three different statistics agree reasonably well about a common value. The peaks of the distributions are not centred precisely about the expected values (black vertical lines) though. The weightings applied to each model to be combined greatly influences the final results. In this instance, models with a mass of 1.29 M$_{\odot}$ of the same $X_{\rm{init}}$ as the input model were preferentially selected compared to 1.25 M$_{\odot}$ models with the same $X_{\rm{init}}$ and 1.27 M$_{\odot}$ models of different $X_{\rm{init}}$ values. All models and weightings used during the analysis process are exported from the program and can be accessed to understand further which models and combinations are preferred for different stars. This can be used to understand and improve the construction of future grids.

The widths of the distributions are related to the uncertainties determined from each run. The uncertainties related to the single run are representative of the formal uncertainties output by AIMS, those of the best MCMC models are expected to be similar to results of the single (unperturbed) run. The test shows that both sets of uncertainties are very similar. Finally, the combined runs have uncertainties equal to the approximate summation of those of the previous two sets in quadrature, as the concatenation of the runs represents both the formal and random uncertainties.

The magnitude of the uncertainties also depends on the underlying grid. An incomplete grid, with insufficient models and/or frequencies will lead to systematic errors in the model selection. Indeed, AIMS rejects models which do not match the entire observed spectrum. The final output parameters are based upon the selection procedure. Hence, anything affecting the accuracy of the selection will affect the final results. As the performance relies upon the input criteria being accepted by a large number of models, an incomplete grid will increase the number of rejections, reducing the accuracy of AIMS. A simple solution (performed here) is to reduce the number of input frequencies in the data file (e.g. limit range of $\nu$ to $\nu_{\rm{max}} \pm 0.5$ $\nu_{\rm{max}}$), increasing the probability for models to match the input criteria. 

To further examine the quality of the results, the number of standard deviations ($N_{\sigma}$) the output parameter ($x_{\rm{calc}}$) lay from the true value ($x_{\rm{true}}$) was determined using 

\begin{equation}
	N_{\sigma} = \frac{x_{\rm{calc}}-x_{\rm{true}}}{\sigma}.
	\label{equ:nsigma}
\end{equation}

\begin{figure}
	\includegraphics[width=\columnwidth]{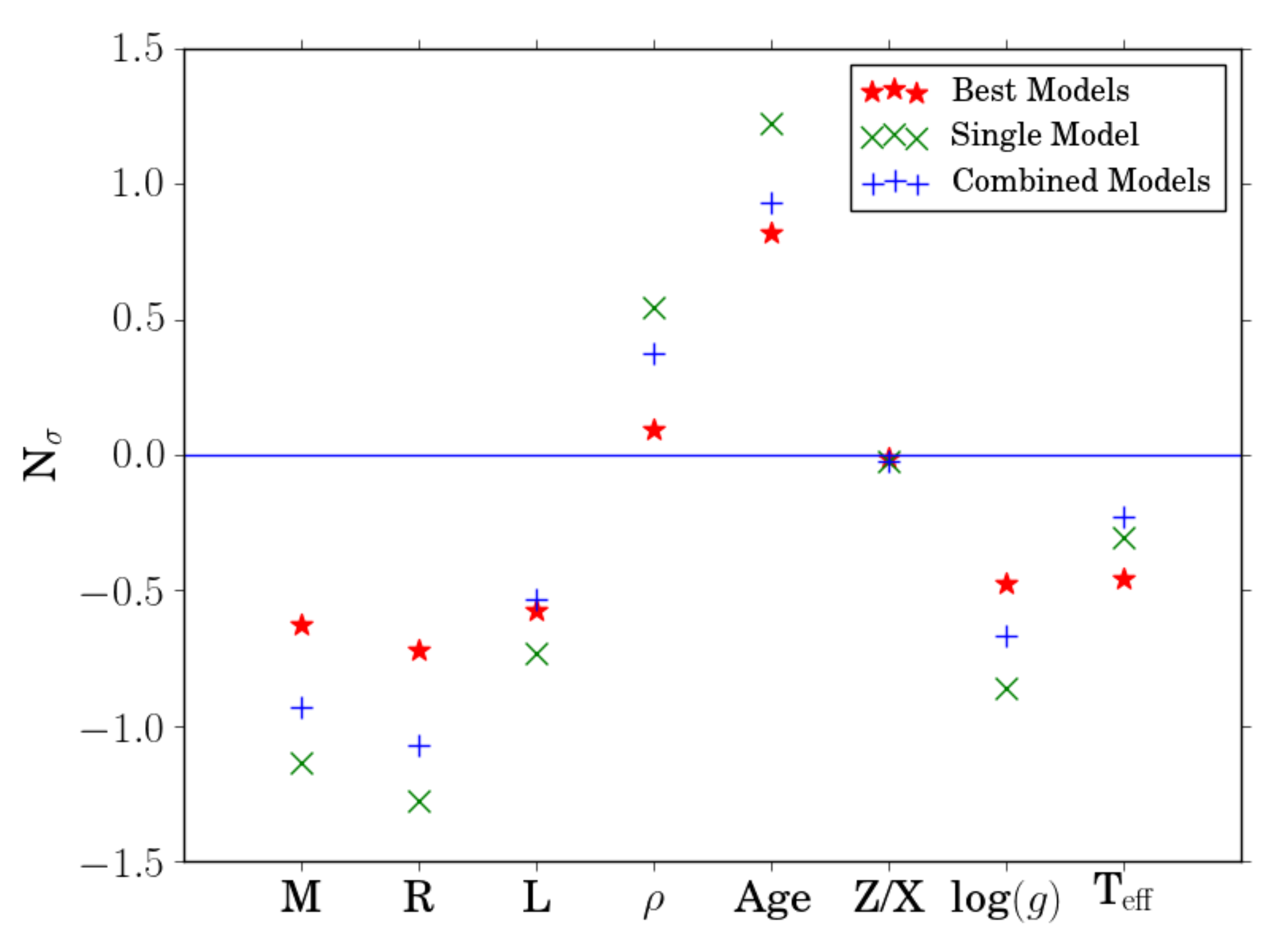}
    \caption{$N_{\sigma}$ steps from the true value of each calculated parameter for the unperturbed (green), combined (blue) and best MCMC (red) models for the MS tests.}
    \label{fig:MS_N-sigma}
\end{figure}

These tests are shown in Fig. \ref{fig:MS_N-sigma}. It is clear that the best MCMC models outperform both combined and single models. The higher performance of the combined run stems from the increased abundance of data, providing a better convergence on the real value than a single run. As the results always lay within $1.5\sigma$ of the result, we can conclude that the fits were successful.

\subsubsection{Red Giants}	\label{RGB_art}

Using the same track and set of classical constraint parameters  for consistency, a model from the RGB grid was selected and subjected to the same tests as the MS model. Frequency uncertainties were constructed as for the MS observational file. Uncertainties on the classical constraints were again consistent with the literature. The period spacing, $\Delta\Pi$, was included as a grid parameter ($\sigma_{\Delta\Pi} = 1\%$, \citealt{2016A&A...588A..87V}) and consequently as one of the outputs in the results. 

\begin{figure*}
    \begin{tabular}{cc}
	\includegraphics[width=\columnwidth]{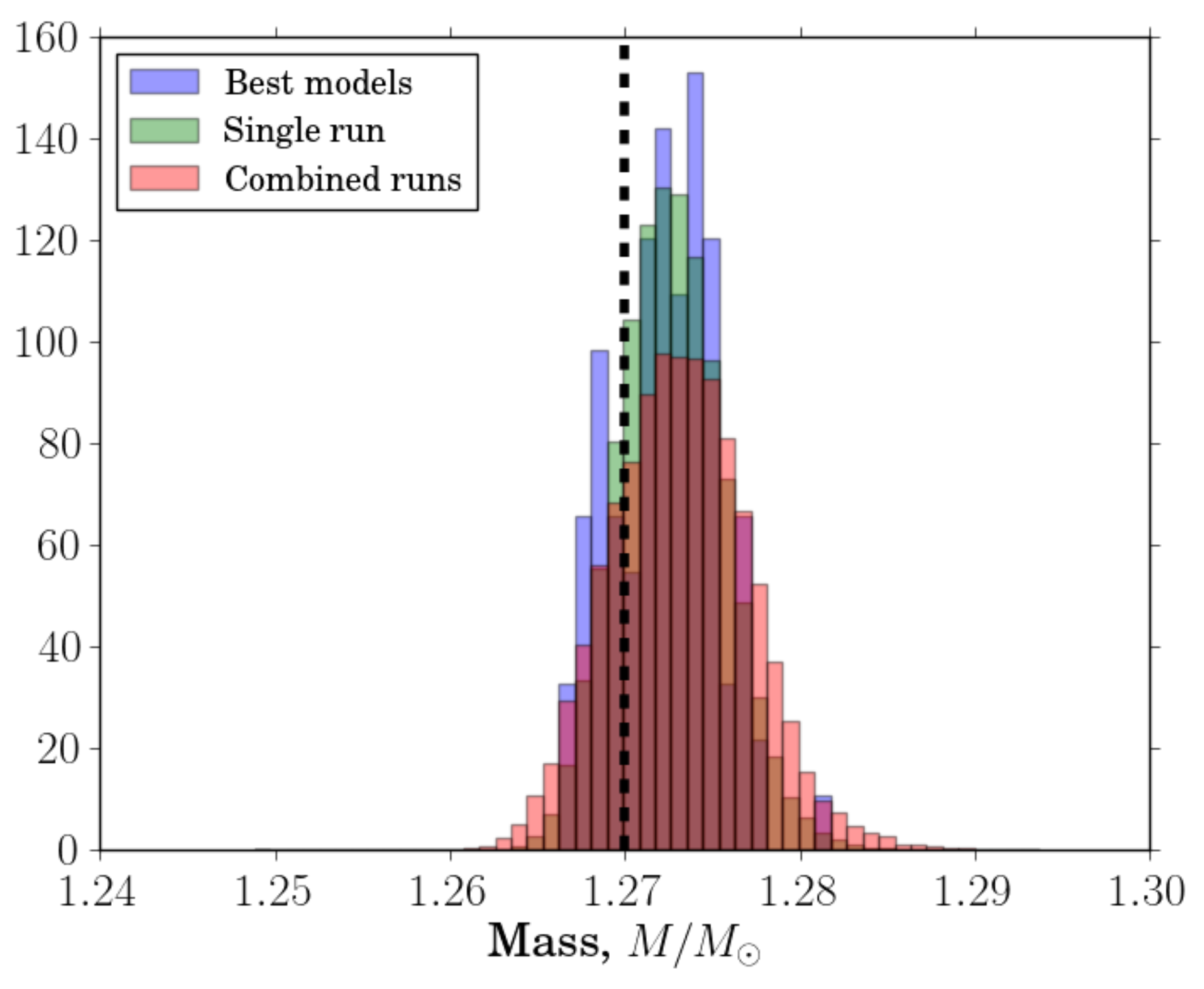} &
	\includegraphics[width=\columnwidth]{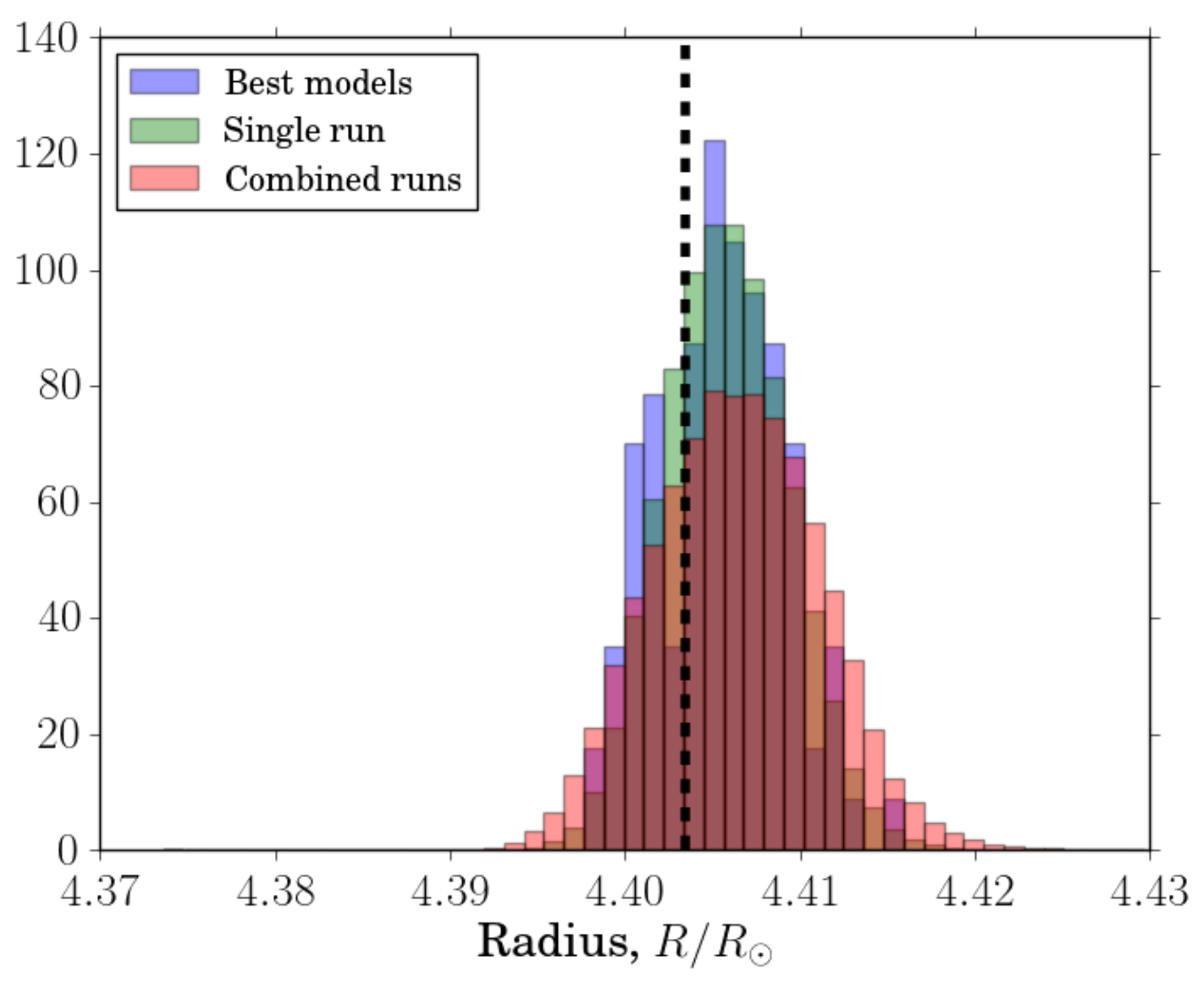}\\
	\end{tabular}
    \caption{The mass (left) and radius (right) PDF distributions for RGB the single model (green), 100 realisations (red) and best models from the MCMC runs (blue). The model mass and radii are 1.27 M$_{\odot}$ and 4.403 R$_{\odot}$, indicated by the vertical black, dotted line.}
    \label{fig:hist_m_RGB}
\end{figure*}

As before, PDFs of the mass and radius, in addition to an $N_{\sigma}$ plot for all parameters, have been included. Figure \ref{fig:hist_m_RGB} shows a tight relationship between each of the three model runs, sharing common peak values. The widths of the distributions of the RGB PDFs are broader than their MS counterparts. This is reflected in the increased uncertainties of the output values. Using fewer frequencies compared to the MS runs (9 RGB, 21 MS) and only $l=0$ modes may contribute to this factor, but it is inherent from broader studies that larger RGB compared to MS uncertainties are to be expected. A larger number of models are also rejected when searching the RGB grid, indicating fewer models are likely to be selected around the desired solution.

\begin{figure}
	\includegraphics[width=\columnwidth]{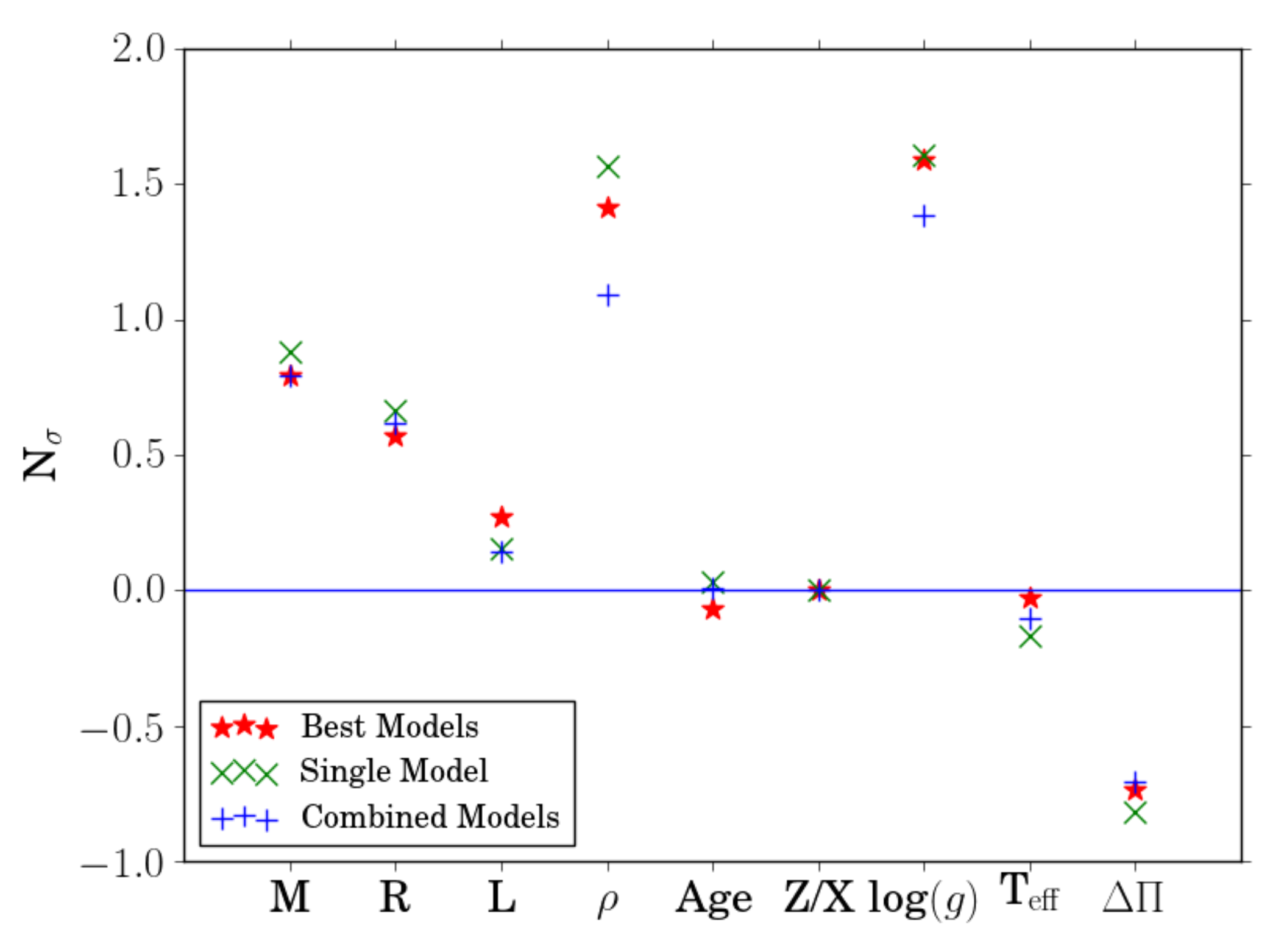}
    \caption{$N_{\sigma}$ steps from the true value of each calculated parameter for the unperturbed (green), combined (blue) and best MCMC (red) models for the RGB tests.}
    \label{fig:RGB_N-sigma}
\end{figure}

The trend in Fig. \ref{fig:RGB_N-sigma} closely resembles that observed in Fig. \ref{fig:MS_N-sigma}, but little should be read into this. Repeating the trials on multiple MS and RGB models from tracks in different regions of the grid resulted in different $N_{\sigma}$ parameter distributions with each track. Each set of parameters returned is subject to different over/under estimations from models resulting from their grid location and the boundary conditions imposed on them. This variation in model determined variables and their associated likelihoods means consistency between $N_{\sigma}$ patterns should not be expected from model to model. The focus should therefore be on the distribution of $N_{\sigma}$ values which are all satisfactorily $< 1.5\sigma$ in each case.

\subsection{The Sun} \label{Real}

Besides artificial data, we used AIMS to reproduce solar data from the BiSON network of telescopes \citep{doi:10.1111/j.1745-3933.2009.00672.x,2014MNRAS.439.2025D,Hale2016}, using the $l = 0, 1, 2$ and $n = 18-23$ modes. The frequency uncertainties were increased by a factor of $\sqrt{21/4}$ \citep{1992ApJ...387..712L,1994A&A...289..649T,2008A&A...486..867B} to perform a Sun-as-a-star analysis. We recall that a solar-calibrated value of the mixing length was used in the grid.

When working with real data, it is necessary to account for surface effects, which are not present in tests performed with artificial data. We used the two-term Ball and Gizon surface correction \citep{ball_new_2014}, although other corrections are also included in AIMS: Ball and Gizon single-term \citep{ball_new_2014}; Kjeldsen \citep{kjeldsen_correcting_2008}; Sonoi (single-term, scaling, two-term - \citealt{sonoi_surface-effect_2015}).

\begin{figure}
	\includegraphics[width=\columnwidth]{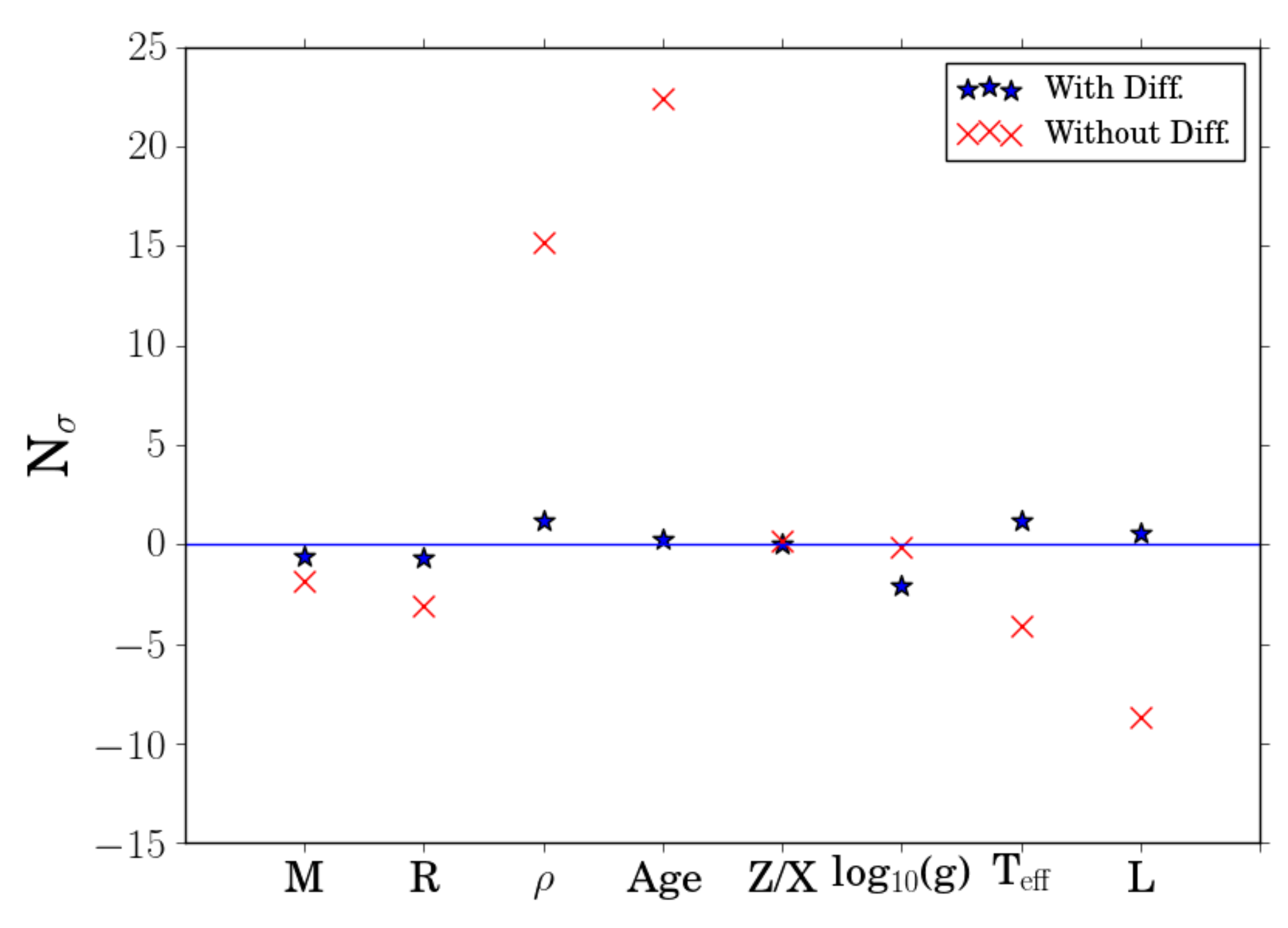}
    \caption{$N_{\sigma}$ steps from the true value of the mass, radius, density, age (taken to be $4.57 \pm 0.02$ Gyr, \citealt{bahcall_solar_1995}), $Z/X$ ratio, log$_{10}$(g), $T_{\rm{eff}}$ and luminosity for the Sun for the grids with (blue stars) and without (red crosses) diffusion.}
    \label{fig:sun_stat}
\end{figure}

The fits were performed using two grids: the nominal CL\'{E}S MS grid and an identical grid, but with microscopic diffusion included in the modelling (re-calibrated mixing length: 1.81). From Table \ref{tab:solar_model_comps}, we can see that models without diffusion can reproduce quite well both the solar mass and radii, although not at the 1$\sigma$ level, but that they present inaccuracies in age of about $\sim$1Gyr. This is in agreement with helioseismic results which reject solar models without microscopic diffusion \citep{1993ApJ...403L..75C}. However, models with microscopic diffusion show excellent agreement with solar values \citep{thoul_element_1994}. Fig. \ref{fig:sun_stat} confirms this, displaying $N_{\sigma}$ results for multiple parameters of the Sun for grids with (blue stars) and without (red crosses) microscopic diffusion.

\begin{table}
	\centering
	\caption{Comparison of Solar parameters using grids with and without microscopic diffusion. Mass and radius are given in Solar units, density in g cm$^{-3}$ and age in Myrs. Literature density and age are from \protect\cite{2012A&A...539A..63R} and \protect\cite{bahcall_solar_1995}.}
	\label{tab:solar_model_comps}
	\begin{tabular}{cccccc} 
		\hline
		Parameter & With Diff. & Without Diff. & Literature\\
		\hline
		Mass & $0.997 \pm 0.005$ & $0.994 \pm 0.003$ & 1.0\\
		Radius & $0.999 \pm 0.002$ & $0.996 \pm 0.001$ & 1.0\\
		$<\rho>$ & $1.412 \pm 0.001$ & $1.4183 \pm 0.0005$ & $1.4104 \pm 0.0012$\\
		Age & $4578 \pm 31$ & $5264 \pm 31$ & $4570 \pm 20$\\
		\hline
	\end{tabular}
\end{table}

\begin{figure*}
    \begin{tabular}{cc}
    With Diffusion & Without Diffusion\\
	\includegraphics[width=0.48\textwidth]{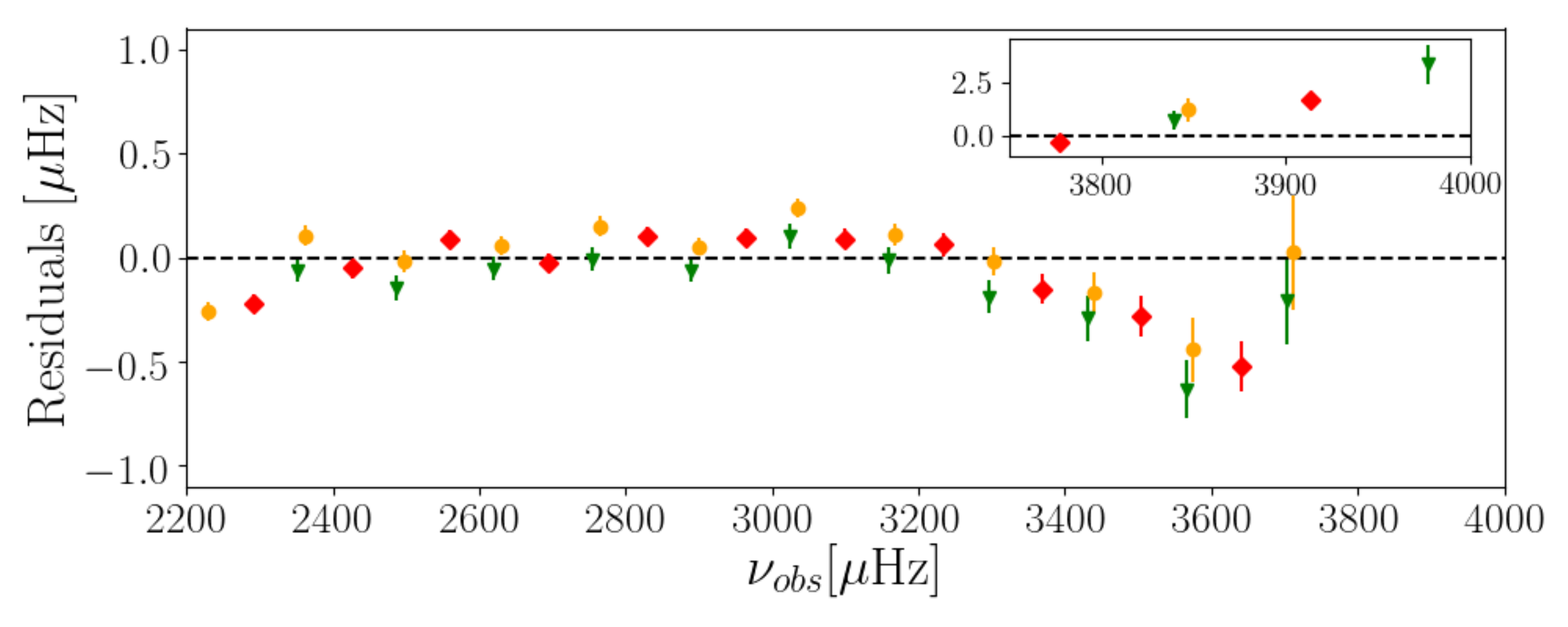} &
	\includegraphics[width=0.48\textwidth]{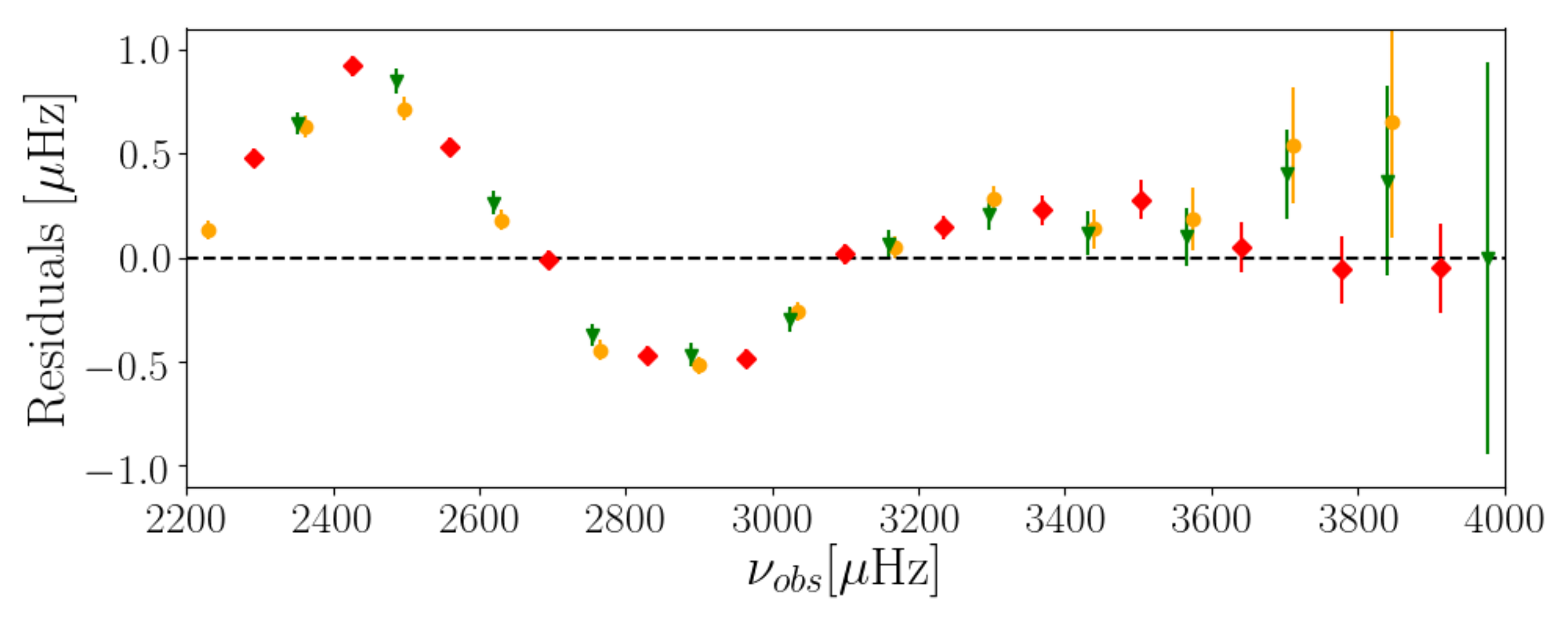}\\
	\end{tabular}
    \caption{Frequency residuals ($\nu_{\rm{obs}}-\nu_{\rm{theo}}$) comparison between the observed and theoretical frequencies output by AIMS for the grid with (left) and without (right) microscopic diffusion. The residuals subplot for the results with diffusion shows the residuals for frequencies $> 3750 \mu$Hz. These residuals are much larger and therefore shown in a separate subplot to allow the underlying trend in the residuals to be observed.}
    \label{fig:sun_w}
\end{figure*}

Figure \ref{fig:sun_w} shows the difference between the observed frequencies ($\nu_{\rm{obs}}$) and the theoretical (surface corrected - s.c., $\nu_{\rm{theo\,s.c.}}$) frequencies returned by AIMS for the grids with (left) and without (right) diffusion respectively. All available Solar frequencies are shown. Residuals are shown to illustrate the quality of the interpolation process, hence the robustness of the parameter determinations. A periodic trend is seen in both cases, 
with a much higher emphasis for the non-diffusive grid. This trend is the result of the large mismatch of helium abundance between the theoretical model and the Sun. Larger disparities are also observed above 3700 $\mu$Hz as a consequence of the surface effects. This clearly illustrates the difficulties and weaknesses of using individual frequencies as direct constraints as the surface effects could bias the modelling results. Using constraints such as frequency ratios for MS stars can help mitigate such effects.

Looking at the reduced $\chi^{2}$ values of the frequencies, it seems sensible to favour the values of the Solar parameters determined using the grid including diffusion:\\ 

\begin{equation}
\chi^{2} = \sum_{\rm{i}} \frac{(\nu_{\rm{theo\,s.c.,i}} - \nu_{\rm{obs,i}})^{2}}{\sigma_{\rm{i}}^{2}}
\end{equation}

\begin{equation}
\chi_{\rm{red}}^{2} = \frac{\chi^{2}}{d.o.f.}
\end{equation}

where $d.o.f.$ (degrees of freedom) is the number of input parameters minus the number of free parameters, and $\sigma$ the compound uncertainty on the frequencies. Indeed, $\chi^{2}_{\rm{red,diff}} = 11.4$ whereas  $\chi^{2}_{\rm{red,non-diff}} = 52.5$. Testing the hypothesis that the model values are true, p values of 0.88 ($\chi^{2}_{\rm{red,diff}}$) and 0.07 ($\chi^{2}_{\rm{red,non-diff}}$) were returned. The order of magnitude difference between the reduced $\chi^{2}$ values clearly indicates that the grid including diffusion is superior for the Solar analysis. As mass, radius and age are all within 1$\sigma$ of the Solar values for this grid, we can conclude that the processes within AIMS perform well enough to produce the results to a high degree of accuracy. All of the uncertainties are lower than one would expect to find in the literature (see \citealt{2017ApJ...835..173S} for recent Solar values from multiple grids and codes) as they are of the same order of magnitude as in tests using artificial data.

\begin{table}
	\centering
	\caption{Solar parameters and uncertainties determined by AIMS using the frequency separation ratios r$_{0,1}$, r$_{0,2}$ and r$_{1,0}$ as asteroseismic constraints. Mass and radius are given in Solar units, density in g cm$^{-3}$ and age in Myr. The diffusive grid was used.}
	\label{tab:solar_uncerts}
	\begin{tabular}{cccccc} 
		\hline
		Parameter & r$_{0,1}$ & r$_{1,0}$ & r$_{2,0}$\\
		\hline
		Mass & $1.01 \pm 0.02$ & $1.00 \pm 0.02$ & $0.99 \pm 0.03$\\
		Radius & $1.01 \pm 0.03$ & $1.00 \pm 0.03$ & $0.99 \pm 0.03$\\
		$\rho$ & $1.40 \pm 0.10$ & $1.42 \pm 0.10$ & $1.44 \pm 0.10$\\
		Age & $4614 \pm 258$ & $4549 \pm 204$ & $4603 \pm 139$\\
		\hline
	\end{tabular}
\end{table}

To determine whether the small uncertainties resulted from the model or the use of individual mode frequencies, the Solar data was also tested using the r$_{0,1}$, r$_{0,2}$ and r$_{1,0}$ frequency separation ratios (\citealt{2003A&A...411..215R}; grid including microscopic diffusion used). An improvement in the returned parameters can be expected, as the ratios focus more on the stellar interior \citep{2003A&A...411..215R,2005MNRAS.356..671O}. Additionally, their reduced sensitivity to surface effects should also lead to an improvement. This is confirmed by the results in Table \ref{tab:solar_uncerts}. The frequency ratios give values consistent with the $\nu_{\rm{ind}}$ results and the expected Solar values, but with larger uncertainties. When using solely the frequency ratios\footnote{AIMS allows the use of other constraints along frequency ratios, such as the large frequency separation, while self consistently keeping track of the correlations between seismic indicators.}, one filters out additional information (e.g. on the mean density of the star) and thus naturally the uncertainties are increased. While this leads to larger uncertainties on the stellar parameters, this degree of precision can also be seen as more robust with respect to systematic effects which can be underestimated when directly fitting the individual frequencies.

\section{Impact of using different combinations of seismic and non-seismic constraints} \label{obs_cons}

In addition to testing the main functionalities of AIMS, the effect of the inclusion of certain combinations of constraints within the input observation file were explored. For all tests presented so far, the classical constraints used have been $\nu_{\rm{max}}$, $T_{\rm{eff}}$, $L$ and [Fe/H]. In addition to these constraints, equal weighting has been given to the asteroseismic (input frequencies, $\nu_{i}$) and classical constraints.

\subsection{Main Sequence Fits} \label{MSF}

Four tests were performed on a single main sequence model (the same model used in Sect. \ref{MS_art}) from within the grid, with the effect on the PDF distributions and uncertainties of the mass, radius and age of the artificial star recorded. The constraint variations were as follows:

\begin{enumerate}
	\item $T_{\rm{eff}}, L (\sigma_{Gaia})$, [Fe/H], no acoustic oscillation frequencies
	\item $T_{\rm{eff}}$, [Fe/H], acoustic oscillation frequencies, no $L$
    \item $T_{\rm{eff}}, L (\sigma_{Gaia})$, [Fe/H], r$_{0,2}$
    \item $T_{\rm{eff}}, L (\sigma_{Gaia})$, [Fe/H], acoustic oscillation frequencies
\end{enumerate}

\begin{figure*}
    \includegraphics[height=6cm,width=\textwidth,keepaspectratio]{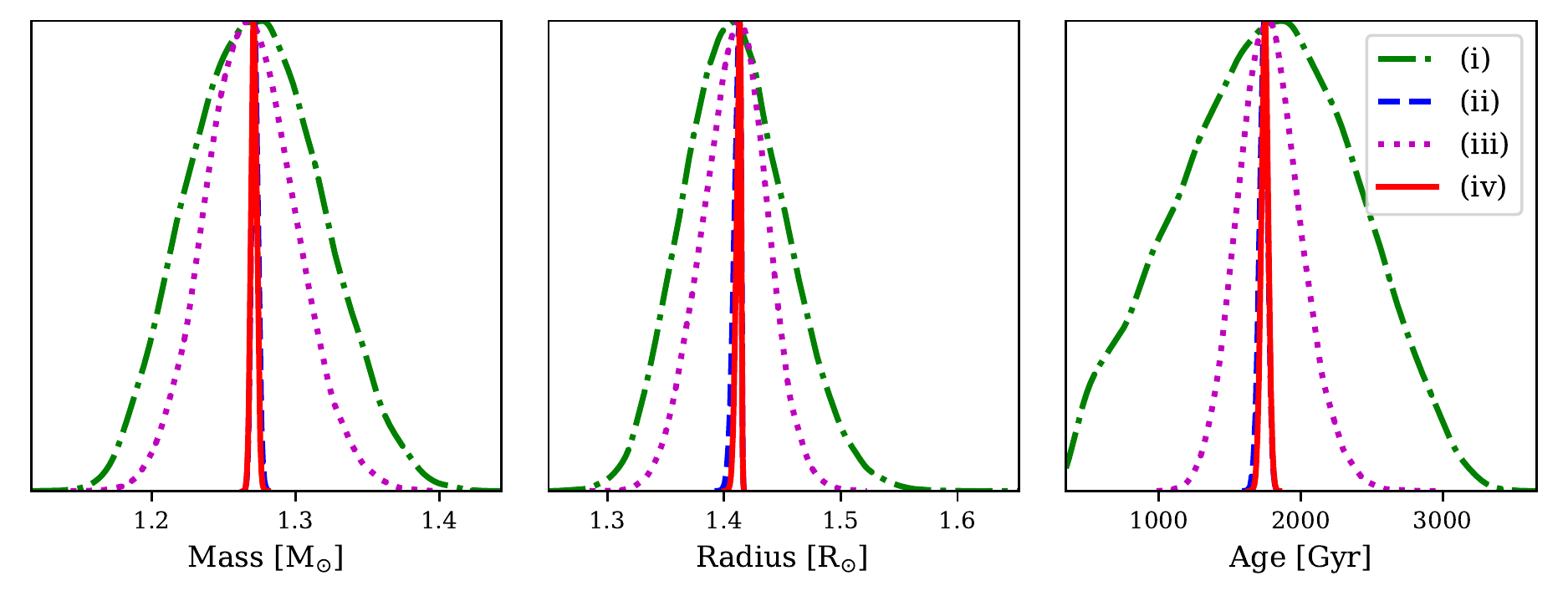}\\
    \caption{Comparison of input classical and seismic constraints for mass (left), radius (centre) and  age (right) determination. The normalised distributions represent the different classical constraint criteria: (i) - green, dot-dashed; (ii) - blue, dashed; (iii) magenta, dotted; (iv) - red, solid.}
    \label{fig:Age_comp}
\end{figure*}

Figure \ref{fig:Age_comp} displays the PDFs for mass, radius and age determinations. The inclusion or exclusion of luminosity from the constraints appears to have a minimal impact on the precision between cases (ii) and (iv). The increase in precision on each parameter may not be as significant as the inclusion of asteroseismology (narrowing of distributions observed in both cases), but an improvement is still observed when luminosity is included. Should other parameters (e.g. initial He abundance) be free to vary, an independent constraint on luminosity is important to lift any existing degeneracies present when using only seismology.

Table \ref{tab:constraint_tests} shows the uncertainties for each set of constraints. Test (i) returns uncertainties of order of the typical literature values for age, mass and radii respectively, but tests (ii) and (iv) return values at least an order of magnitude smaller. The addition of more free parameters to the grid and the intrinsic differences they would cause between models would increase these uncertainties to be closer to those expected. However, the trend between constraint sets is clear. The decision to include or exclude the acoustic oscillation frequencies has a significant impact on all parameters, reducing the percentage errors by an order of magnitude.

Though reduced compared to case (i), case (iii) uncertainties are of the same order of magnitude despite the inclusion of asteroseismology and are in line with the best literature values. This illustrates the difference in precision achievable with the inclusion of global asteroseismic parameters compared to the use of individual mode frequencies when the same classical constraints are available. The potential improvement in precision to be gained underlines the importance of the development of analysis codes, such as AIMS, capable of using individual acoustic oscillation frequencies for the furthering of asteroseismic studies.

\begin{table}
	\centering
	\caption{Percentage uncertainties for the determined values of mass, radius and age for the MS model used in the observational tests, subject to the tested combinations of classical and asteroseismic constraints.}
	\label{tab:constraint_tests}
	\begin{tabular}{ccccc} 
		\hline
		Constraint & Age ($\%$) & Mass ($\%$) & Radius ($\%$)\\
		\hline
		(i) & 34.64 & 3.69 & 3.12\\
		(ii) & 1.69 & 0.18 & 0.22\\
		(iii) & 12.65 & 2.48 & 1.98\\
		(iv) & 1.41 & 0.16 & 0.18\\
        \hline
	\end{tabular}
\end{table}

\subsection{Red Giant Fits}

The process was repeated for an RGB model from the grid to illustrate that, despite less convincing interpolation results than on the MS, it also performs well in this regime. 
Consequently, a more comprehensive approach was taken. We compare the results of AIMS for red giant stars to an extensively used, pre-existing stellar parameter determination code to prove the capability of AIMS as an analysis tool. We chose the PARAM software \citep{2006A&A...458..609D,2014MNRAS.445.2758R,2017MNRAS.467.1433R}, which is quite similar to AIMS in its philosophy, with the only significant difference being that AIMS uses asteroseismic data as an input. We run AIMS without using the input mode frequencies to make a more informed comparison between the capabilities of both codes.

A recent work by \cite{2017MNRAS.467.1433R} (hereafter R17) investigates the effects of various combinations of constraints on the accuracy of stellar parameter determinations for a series of artificial red giant and red clump stars using PARAM. We repeated these tests using the same sets of classical and global seismic constraints in AIMS and our own RGB model. 10 different combinations of constraints were used:

\begin{flushleft}
\begin{tabular}{rlccrl}
	(i) & $\Delta\nu$ & & & (vi) & $\Delta\nu, \nu_{\rm{max}}, \Delta\Pi, L$\\
	(ii) & $\Delta\nu, \nu_{\rm{max}}$ & & & (vii) & $\nu_{\rm{max}}, L$\\
	(iii) & $\Delta\nu, \Delta\Pi$ & & & (viii) & $\log_{10}(g), L$\\
	(iv) & $\Delta\nu, \nu_{\rm{max}}, \Delta\Pi$ & & & (ix) & $\Delta\nu, \log_{10}(g)$\\
	(v) & $\Delta\nu, \Delta\Pi, L$ & & & (x) & $\Delta\nu, L$\\
\end{tabular}
\end{flushleft}

From the above list, it is clear that asteroseismic parameters are still to be used as initial constraints with the large frequency separation ($\Delta\nu, \sigma_{\Delta\nu} = 0.05 \mu$Hz), frequency of maximum power ($\nu_{\rm{max}}, \sigma_{\nu_{\rm{max}}} = 2\%$) and period spacing ($\Delta\Pi, \sigma_{\Delta\Pi} = 1\%$) featuring heavily. These parameters are all global seismic properties and are not necessarily reliant upon determination of individual frequencies. Hence, they can be input as classical constraints. In addition to the listed constraints, the effective temperature ($\sigma_{T_{\rm{eff}}} = 80$ K) and metallicity ($\sigma_{[\rm{Fe/H}]} = 0.1$ dex) were included for each case as in R17. The uncertainties used on $L$ and log$_{10}$(g) were $3\%$ and 0.1 dex respectively. It should be noted that case (iii) of R17 was ignored here, with a value of $\Delta\nu$ calculated from the frequencies used throughout.

\begin{figure*}
	\includegraphics[width=0.95\textwidth,keepaspectratio]{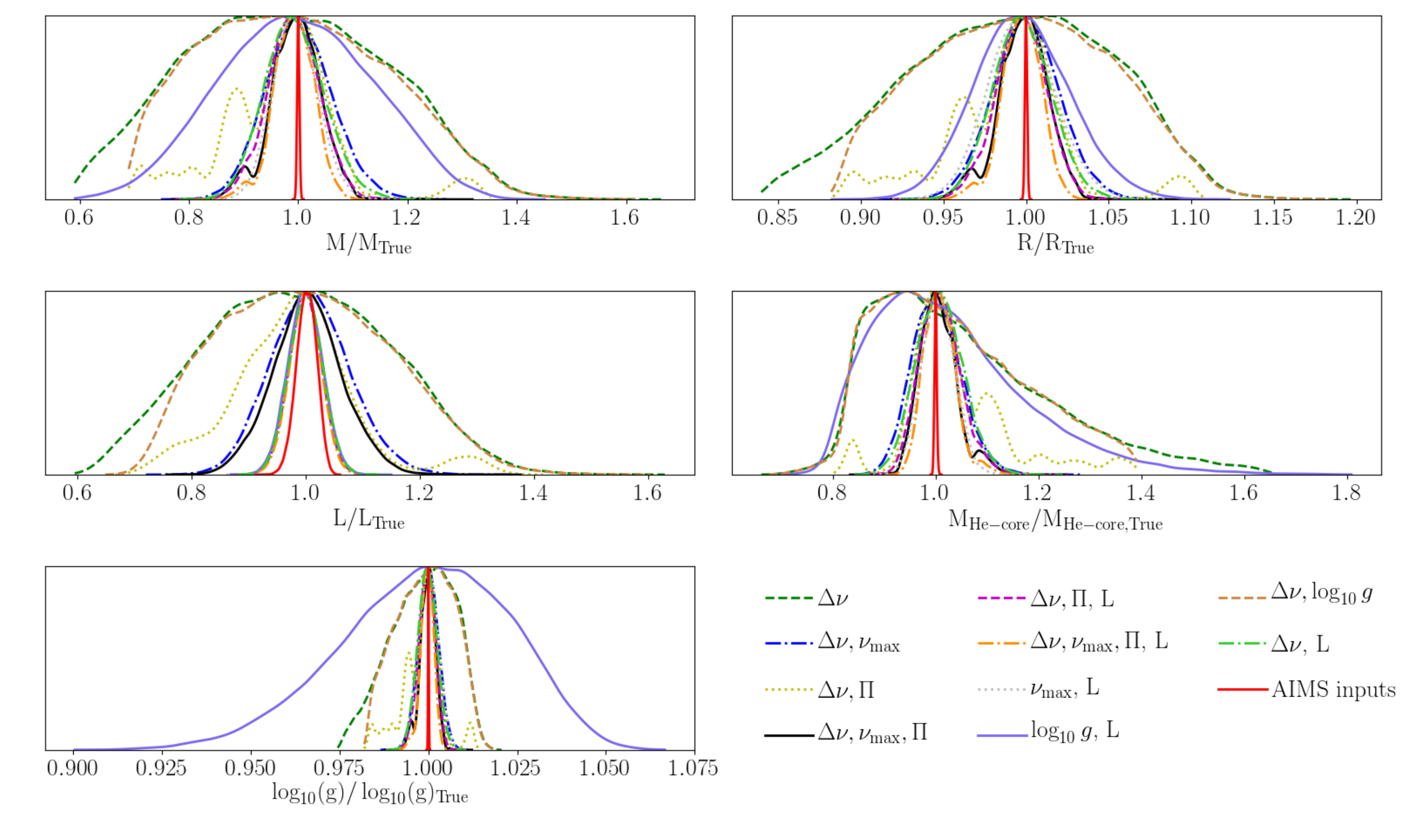}
    \caption{Comparison of the posterior probability distributions for multiple combinations of constraints used as inputs to AIMS without the use of the individual mode frequencies. The distribution marked `AIMS inputs' shows the result obtained if the individual mode frequencies are used.}
    \label{fig:PARAM_comps}
\end{figure*}

Figure \ref{fig:PARAM_comps} displays the results of these tests as the distributions of the determined values normalised to the true parameter values. In addition to the above sets of constraints, the model was tested using the standard constraints used throughout this work and with a direct fit of the individual mode frequencies. This is labelled `$\nu_{i}$'.

To further analyse the distributions, Table 3 from R17 has been recreated. Table \ref{tab:PARAM_tab} contains the relative uncertainties for the mass and age of the tested model for each combination of constraints. The majority of the results follow typically Gaussian distributions, but cases (i), (viii) and (ix) show asymmetry in their mass distributions. The sampling of the mass in these cases has reached the lower end of the grid, introducing a sampling bias as a build up of low mass samples occurs. This causes the asymmetry observed, which propagates to other parameters.

A direct comparison between the two sets of results is not appropriate due to the different models used, but a comparison of the overall trends is meaningful. The attained values vary, but in general the distributions follow those of R17. This is reassuring and confirms that AIMS reacts to certain combinations of constraints in an expected manner.

\begin{table}
	\centering
	\caption{Fractional uncertainties for each combination of input constraints for AIMS run as PARAM. The RGB results from table 3 of R17 are displayed for direct comparison.}
	\label{tab:PARAM_tab}
	\begin{tabular}{cccccc} 
		\hline
		\multirow{2}{*}{Constraints} & \multicolumn{2}{|c|}{$\sigma_{M}/M$} & \multicolumn{2}{|c|}{$\sigma_{\rm{Age}}/\rm{Age}$}\\
        & AIMS & R17 & AIMS & R17\\
		\hline
		$\nu_{i}$ & 0.002 & - & 0.029 & -\\
		$\Delta\nu$ & 0.184 & 0.173 & 0.735 & 0.734\\
	    $\Delta\nu, \nu_{\rm{max}}$ & 0.061 & 0.078 & 0.230 & 0.284\\
	    $\Delta\nu, \Delta\Pi$ & 0.119 & 0.109 & 0.475 & 0.336\\
	    $\Delta\nu, \nu_{\rm{max}}, \Delta\Pi$ & 0.047 & 0.054 & 0.190 & 0.192\\
	    $\Delta\nu, \Delta\Pi, L$ & 0.048 & 0.043 & 0.131 & 0.122\\
	    $\Delta\nu, \nu_{\rm{max}}, \Delta\Pi$, L & 0.037 & 0.034 & 0.110 & 0.097\\
	    $\nu_{\rm{max}}, L$ & 0.041 & 0.039 & 0.108 & 0.107\\
	    $\log_{10}(g), L$ & 0.138 & 0.124 & 0.544 & 0.427\\
	    $\Delta\nu, \log_{10}(g)$ & 0.166 & 0.173 & 0.590 & 0.727\\
	    $\Delta\nu, L$ & 0.055 & 0.052 & 0.146 & 0.143\\
		\hline
	\end{tabular}
\end{table}

Considering previous statements regarding AIMS uncertainties, the fractional uncertainties shown in Table \ref{tab:PARAM_tab} are comparable to those of PARAM. Removing the use of individual mode frequencies causes the inflation of the uncertainties due to the smaller number of constraints. Consistency between codes here is important to show that when global asteroseismic parameters are used as constraints, AIMS performs as well as a pre-established and trusted software. Some variation of the fractional uncertainties compared to R17 is present but likely stems from the differences in model parameters and grid properties, as well as modelling codes used in these tests.

Using all of the available information from the mode frequencies improved the fractional uncertainties with values of 0.002 and 0.029 in mass and age respectively. This test case also produces the best PDFs in Fig. \ref{fig:PARAM_comps}, showing the potential of using constraints determined from individual frequencies.

In order to demonstrate the difference in performance between using all the available modes and only the global asteroseismic parameters on an MS star, the model used in section \ref{MS_art} was re-run using the same constraints and configuration as test (ii). Though it was not possible to perform such a comparison with PARAM results, the consistency of the AIMS results without the use of individual frequencies with PARAM allows meaningful comparisons.

\begin{figure*}
	\includegraphics[width=0.95\textwidth,keepaspectratio]{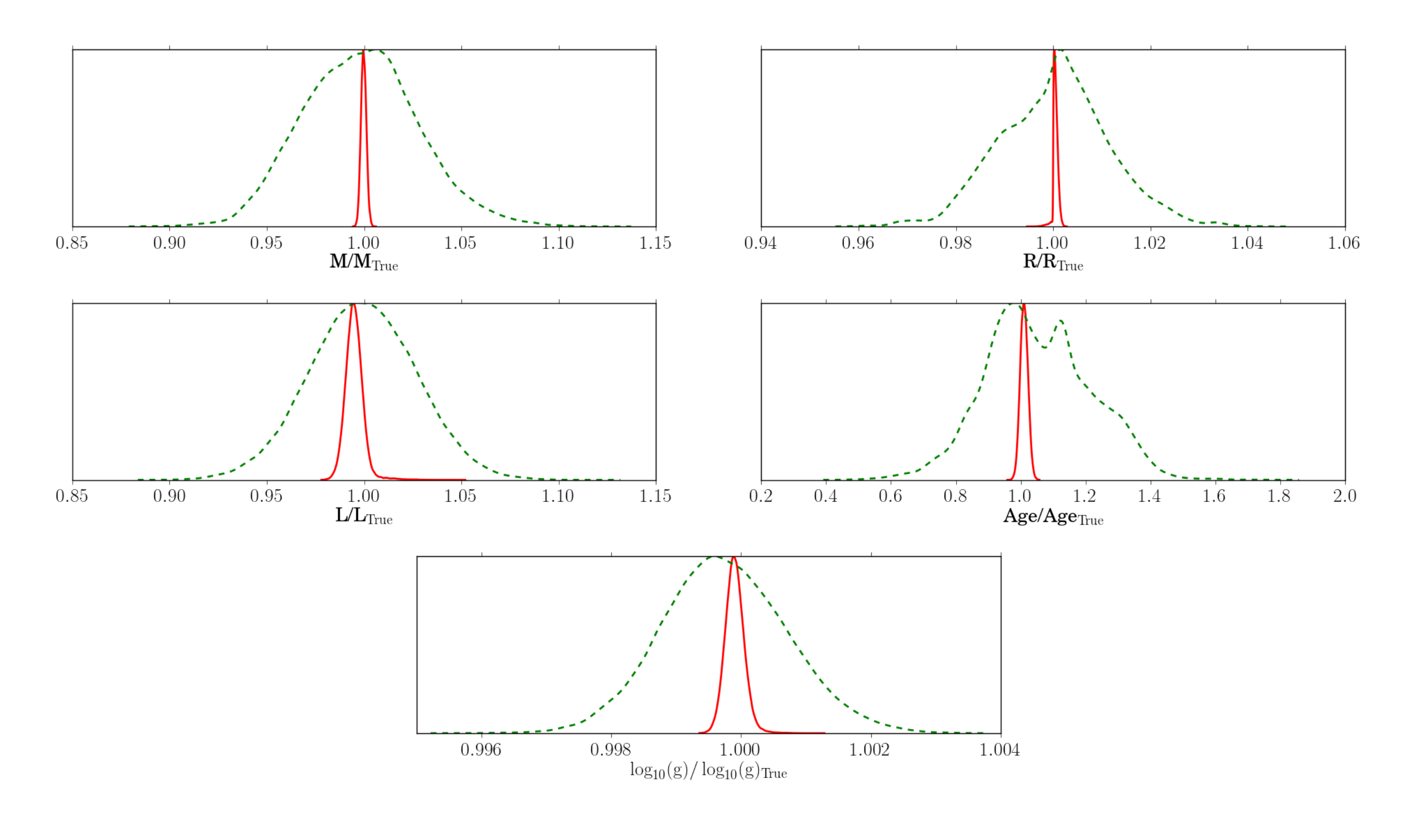}
    \caption{Comparison of the normalised posterior probability distributions for the MS model both with (red, solid) and without (green, dashed) the use of the individual mode frequencies.}
    \label{fig:MS_nu_dnu}
\end{figure*}

Figure \ref{fig:MS_nu_dnu} shows the result comparison of two separate runs for mass and radius as before, as well as the relations for the luminosity, surface gravity and evolutionary parameter - age. An offset between the peaks of the distributions is present for various parameters, caused by the known helium-mass degeneracy (see \citealt{2012A&A...538A..73B} and references therein). As tighter constraints are placed on the luminosity of the star when asteroseismology is used, these degeneracies become lifted, allowing for tighter distributions around the expected solutions.

\begin{table}
	\centering
	\caption{Percentage uncertainty of calculated variables with and without the use of individual frequencies as a constraint for an MS artificial model. `With $\nu$' indicates that the individual frequencies were used in the analysis. $<\Delta\nu>$ indicates the runs without the use of the individual frequencies, but inclusion of the average large frequency separation as a constraint. The $l=0,1,2$ modes were used.}
	\label{tab:nu_dnu_comp}
	\begin{tabular}{ccccccc} 
		\hline
		Model & $\sigma_{\rm{Mass}}$ & $\sigma_{\rm{Radius}}$ & $\sigma_{\rm{Lum}}$ & $\sigma_{\rm{g}}$ & $\sigma_{\rm{Age}}$\\
		\hline
		MS (with $\nu$) & $0.14\%$ & $0.06\%$ & $0.48\%$ & $0.01\%$ & $1.18\%$\\
		MS ($<\Delta\nu>$) & $2.96\%$ & $1.12\%$ & $2.71\%$ & $0.10\%$ & $15.48\%$\\
		\hline
	\end{tabular}
\end{table}

Table \ref{tab:nu_dnu_comp} shows the statistical trends observed in the related figures, giving the percentage uncertainty on each of the relevant parameters for the cases where the individual frequencies were used and when only global asteroseismic constraints were applied.
It should be stressed that the uncertainties displayed in Table \ref{tab:nu_dnu_comp} are purely statistical and do not account for any systematics within the code. Despite this, the effect of including the individual frequencies in the analysis is clear from Table \ref{tab:nu_dnu_comp}. A reduction in uncertainty is seen for all parameters when using the individual frequencies, displaying the benefits of using this additional information. However, as discussed before, the direct use of individual frequencies as constraints can lead to underestimated uncertainties. This is particularly true for main-sequence stars, which often have very rich oscillation spectra. In that sense, being able to use frequency ratios in AIMS allows us to obtain a more realistic precision on stellar parameters and should generally be preferred.

The reduction between the RGB and MS uncertainties is not of the same order for case (vi) for mass and age (case vi contains the initial artificial RGB test classical constraints and therefore is most appropriate to compare). The uncertainties in mass and age on the RGB decrease by factors of 18 and 4, while on the MS the reduction is by a factor of 20 and 15 respectively. The reduction in mass is quite similar for each evolutionary state, but the reduction in age is an order of magnitude greater for the MS. This is mainly due to the observed frequencies and the additional information they carry on the internal stellar structure. Indeed, the RGB fit used 9 frequencies while 21 frequencies were used on the MS. Besides the number differences, the MS fit used modes of $l=0,1$ and $2$, containing a lot of information on the evolutionary stage on the MS, whereas the RGB fit only used radial modes. The $\ell=1$ and $\ell=2$ modes can of course be included within the RGB grid. However, their highly non-linear behaviour and the decreased sensitivity of the small separations to age in evolved stars (e.g. see \citealt{montalban_inference_2010}), currently precludes their direct use in AIMS.

To further illustrate the impact of the inclusion of the $\ell=1$ and $\ell=2$ modes, an additional test on the MS using only the $\ell=0$ modes from the previous tests (7 in total) was performed. The reduction in uncertainty in this instance was only a factor of $\sim 5$ in age, much more in line with the RGB results. This demonstrates the reliability with which stellar parameters can be derived and also points towards the potential improvement to make with RGB grids and data sets containing more than just the $\ell=0$ mode frequencies.

\section{Discussion and Conclusion} \label{Discussion}

We have presented a new, open source, code for the determination of stellar parameters. It is unique as it is currently the sole code using a Bayesian and MCMC algorithm approach with grid interpolation to carry out asteroseismic inferences. The code's flexible, multidimensional approach to the analysis allows the user to analyse data as a function of 2 or more fixed grid parameters, affording more control over the analysis dimensions. We executed a comprehensive testing phase and presented the results. All aspects of the program were analysed, with the results proving satisfactory.

A test of the interpolation procedures revealed the accuracy to which the interpolation function within the program returns known values from within the grid. Primarily, the tests focused on the interpolation of the radial mode frequencies of the MS and RGB grids, showing that AIMS provides accurate interpolations well above the threshold values set from the literature. Additional inputs (e.g. mass, age, radius...) were then also tested and again found to be returned at a level matching or exceeding the desired threshold.

The parameter determination tests were very informative. Primary tests with artificial data shed light on some potential limitations of both the analysis code and underlying grid. We showed that the parameter uncertainties determined by AIMS are approximately an order of magnitude smaller than typically reported in the literature. Further investigations confirmed that the statistical analysis and propagation of observational uncertainties were robust. The uncertainties stated by AIMS are thus statistical and do not account for biases in the input physics or model selection.

The artificial data tests were satisfactory, with parameters lying within a few $\sigma$ of the true results. The input parameters of the model were not returned, but the results were sufficiently close to the input values for this not to be of great concern. Data for the Sun were analysed to test AIMS on real data with clearly defined parameters for comparison. As expected, the precision achieved when including individual oscillation modes leads to a comparable accuracy with the known values only if one has flawless models. As shown here, this objective is yet to be achieved. One can use the evidence from comparisons of a diffusive and non-diffusive grid to highlight the limitations of certain models and the need to improve upon model selection. As AIMS is strongly coupled to the input grid, its performance depends on the standard of the grid used and the final model selection and returned parameters are ultimately a reflection of this.

The AIMS code is highly flexible in terms of the parameter constraints one can use in the analysis. The code can be operated using individual mode frequencies, frequency ratios or large frequency separations as asteroseismic constraints. It is also possible to operate the code without these options, simply using classical constraints instead. Full posteriors are returned for determined parameters in each case, meaning any correlations are taken care of in the analysis process. The effect of altering the classical and asteroseismic constraints associated with the input observational file was explored, with the impact of including or excluding any asteroseismic constraints extremely clear. The inclusion of asteroseismic constraints improved the internal precision by a factor 2-20 for both the RGB and MS stars respectively for all of the tested parameters ($M, R, \rho$, age, $Z/X$, log$_{10}$(g), $T_{\rm{eff}}$, $L$), underlining how important asteroseismology is to AIMS for accurate inferences and the improvement in measurements this technique allows for. 

A comparison with an established stellar parameter code, PARAM, gave a valuable insight into the performance of AIMS. For red giant stars, a set of artificial data similar to those used in R17 was analysed using a variety of constraint combinations, including multiple global asteroseismic parameters. Some variation from the expected values for different combinations was observed, but upon comparison with the work in R17, the distributions and relative uncertainties show comparable trends. The similarity in results to an established code brings confidence to those being output by AIMS, showing that it performs to the standard expected by the field, even without the use of the individual mode frequencies it is designed to use.

The primary focus of constraint testing was on the precision to which the code can operate, but pushing it to the challenging limits of using the best constraints - i.e. individual oscillation frequencies with uncertainties of the order of $10^{-2} \mu$Hz. The robustness shown here by the results achieved give confidence to explore more possibilities with the code. 

Our tests show that, when using individual mode frequencies as constraints, one is in principle able to infer properties with exceedingly high precision. The latter, however, should not be taken as realistic expectations concerning accuracy. Individual mode frequencies are affected by systematic effects that will dominate the uncertainties on the inferred properties. We do not explore such effects in this work, except from the enlightening case of the Sun, where fitting individual mode frequencies results in very high precision estimates of its global properties, which are, however, highly inaccurate if one uses inaccurate models (see Sec. \ref{Real}). This strong model-dependence is attenuated when one uses frequency ratios, as shown in the literature, at the cost of a reduced precision. 
Explorations of the systematic uncertainties in the models and the inclusion of additional free parameters (e.g. $Y_{\rm{init}}$, mixing length, surface effects) provide additional challenges to progress the code and maintain a high quality analysis tool for the community, and will be presented in a forthcoming work (e.g. see \citealt{2018MNRAS.477.5052N}).
On a positive note, and as demonstrated by the tests using Solar data, AIMS can be used for the comparison of competing models which can be selected using Bayesian inference, as derived from the full posterior distributions of various estimated properties.
These tests will be instrumental to promote the development of next generation stellar models, and will improve our ability to determine stellar ages and chemical yields,  with wide impact e.g. on the characterisation and ensemble studies of exoplanets, on evolutionary population synthesis, integrated colours and thus ages of galaxies.

The overall outcome of this work has proven Asteroseismic Inference on a Massive Scale to be a flexible, high precision stellar parameter determination program, fit for use to bring tighter constraints to the determinations of stellar parameters through robust asteroseismic analysis and grid modelling for both dwarf and giant stars. Its flexibility and open-source nature makes AIMS a suitable starting point for the development of the pipelines of future missions such as PLATO (\citealt{2014ExA....38..249R}). Moreover, its output can also be used for additional seismic investigations with for example non-linear inversion techniques as developed by \citet{2002ESASP.485...75R} or linear inversions of structural indicators \citep{2012A&A...539A..63R,2015A&A...583A..62B,2018A&A...609A..95B}.

\section*{Acknowledgements}

The `Asteroseismic Inference on a Massive Scale' (AIMS) project was developed at the University of Birmingham by Daniel R. Reese as one of the deliverables for the SPACEINN network. The SPACEINN network is funded by the European Community's Seventh Framework Programme (FP7/2007-2013) under grant agreement no. 312844. We gratefully acknowledge the support of the UK Science and Technology Facilities Council (STFC). Funding for the Stellar Astrophysics Centre is provided by the Danish National Research Foundation (Grant DNRF106).  BMR, AM, and GRD are grateful to the International Space Science Institute (ISSI) for support provided to the \mbox{asteroSTEP} ISSI International Team. A.M. and G.B. acknowledge support from the ERC Consolidator Grant funding scheme ({\em project ASTEROCHRONOMETRY}, G.A. n. 772293). T. L. Campante acknowledges support from the European Union's Horizon 2020 research and innovation programme under the Marie Sk\l{}odowska-Curie grant agreement No. 792848 and from grant CIAAUP-12/2018-BPD. BN is supported by Funda\c{c}\~{a}o para a Ci\^{e}ncia e a Tecnologia (FCT, Portugal) under the Grant ID: PD/BD/113744/2015 from PHD:SPACE an FCT PhD program and
through national funds (UID/FIS/04434/2013), and by FEDER - Fundo Europeu de Desenvolvimento Regional funds through the COMPETE 2020 - Operacional Programme for Competitiveness and Internationalisation (POCI), and by Portuguese funds through FCT in the framework of the project POCI-01-0145-FEDER-007672, POCI-01-0145-FEDER-028953, POCI-01-0145-FEDER-030389. We also wish to thank Dr M. Long for his contributions.
\\
\\



\bibliographystyle{mnras}
\bibliography{AIMS}


\appendix
\section{}

\subsection{Track Interpolation} \label{Track_Interp}

\begin{figure*}
	\includegraphics[width=\textwidth,keepaspectratio]{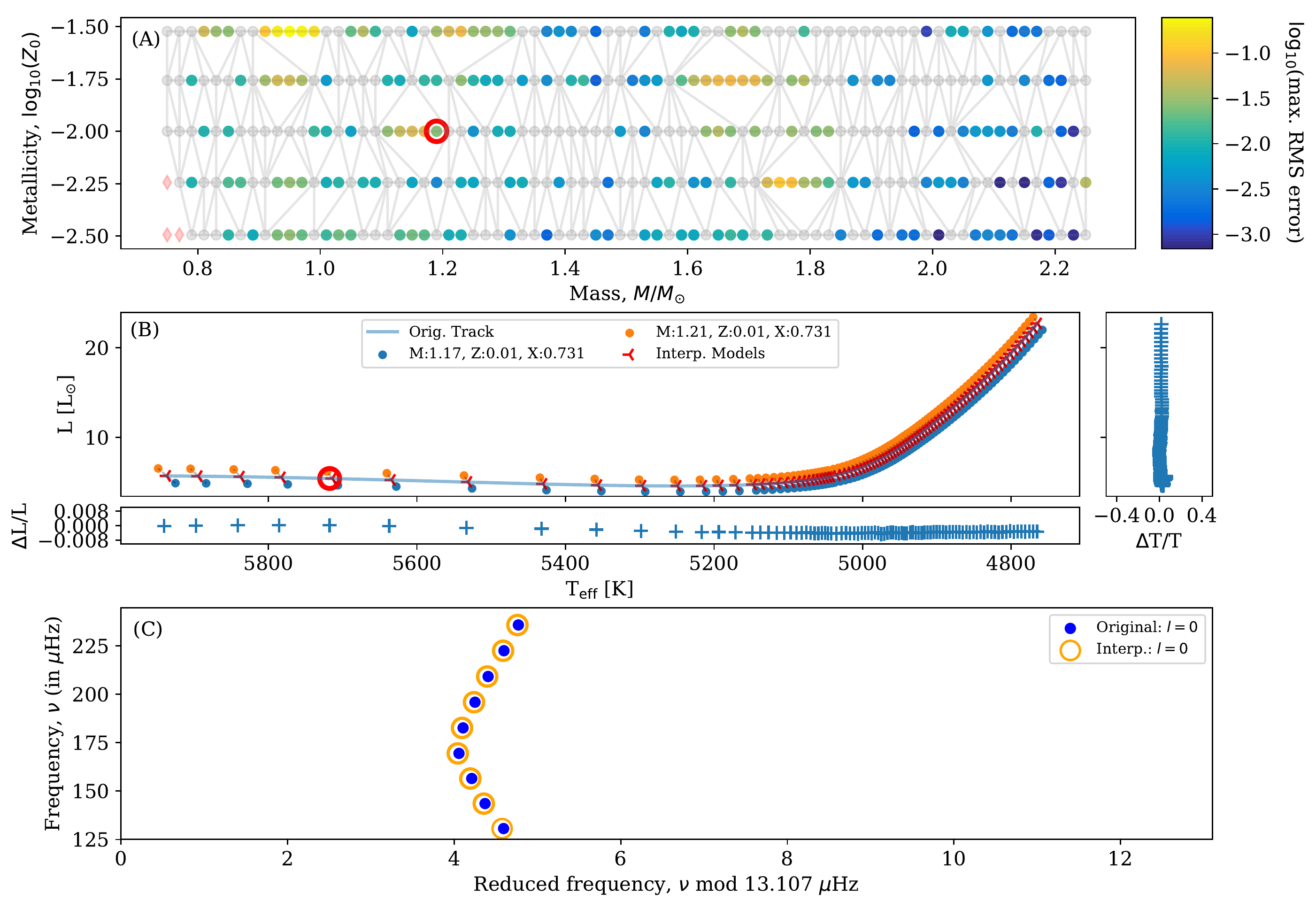}
    \caption{Results as for Fig. \ref{fig:MS_trackI}. A 1.19 M$_{\odot}$, $X_{\rm{init}}$ = 0.731, $Z_{\rm{init}}$ = 0.0100 track is tested here, but with a model from the RGB grid. A maximum interpolated frequency error of -2.088 is returned for this track and the mass of the Helium core is used as the interpolation parameter. The values of $\Delta$T/T have been increased by a factor of 100 for ease of plotting. The frequencies in (C) have been increased by 5$\mu$Hz to centre the frequency pattern.}
    \label{fig:RGB_trackI}
\end{figure*}

Figure \ref{fig:RGB_trackI} is as Fig. \ref{fig:MS_trackI}, but for a track/model selected from the interpolation tests of the RGB grid. A 1.19 M$_{\odot}$, $X_{\rm{init}}$ = 0.731, $Z_{\rm{init}}$ = 0.0100 track is recovered here from models 0.02M$_{\odot}$ either side of the original track and of the same $X_{\rm{init}}$ and $Z_{\rm{init}}$ values. The frequency replication is slightly more uncertain than for the MS example (max. log$_{10}$ error of -2.088), but again excellent parameter residuals and minimal shifts in frequencies show that the process is working well.

\subsection{Parameter Interpolation} \label{param_interp}

\begin{figure*}
\includegraphics[width=0.48\textwidth]{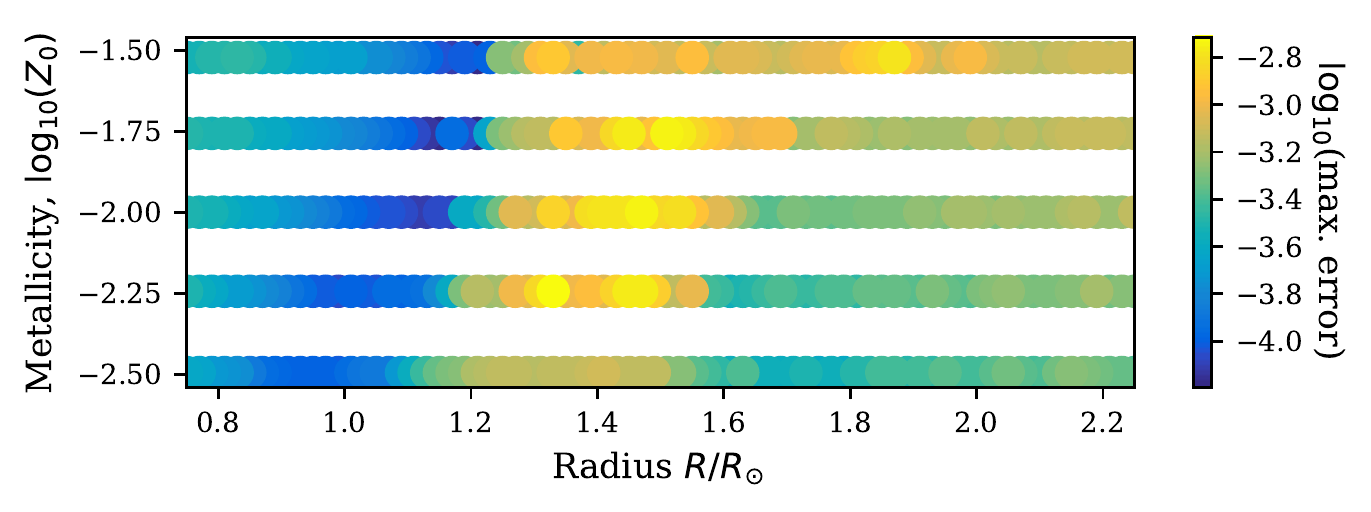}
\includegraphics[width=0.48\textwidth]{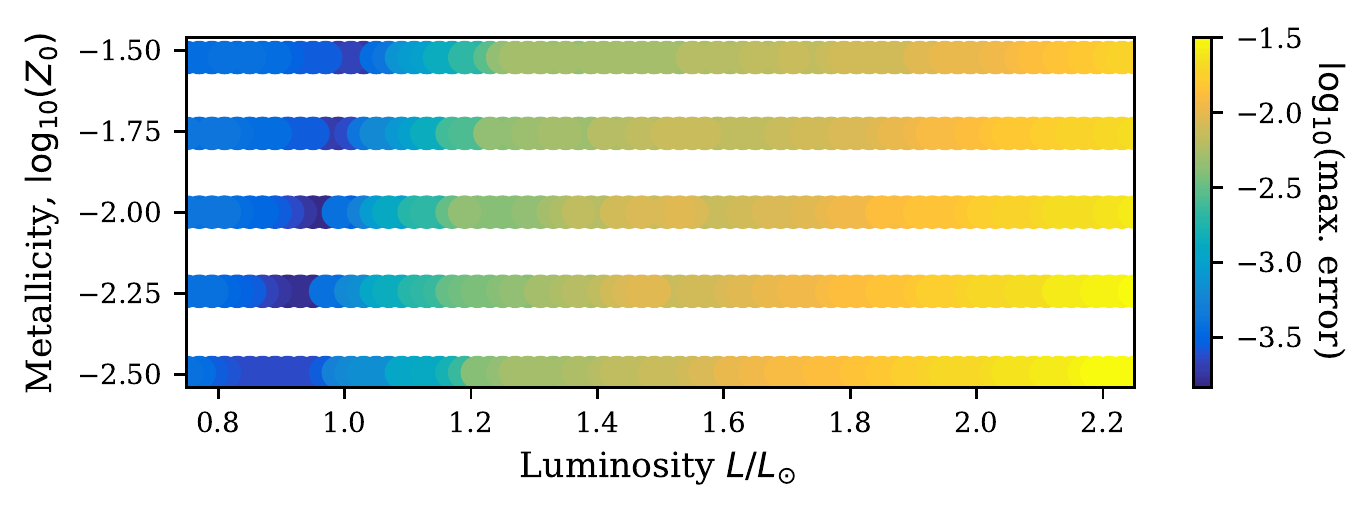}
\caption{\textit{Left}: Magnitude of the radius errors for the MS grid. \textit{Right}: Magnitude of the luminosity errors for the MS grid. Interpolation from grid points a single increment from the original solution. The black circles show the grid node points. Uncertainties in Solar units.}
\label{fig:interp_Lum}
\end{figure*}

Examples of the interpolation plots for radius and luminosity for the artificial main sequence star analysed in the main text are shown in Fig. \ref{fig:interp_Lum}. The maximum uncertainty for each evolutionary track is shown as per the main text.

\bsp	
\label{lastpage}
\end{document}